\documentclass[prd,,aps,showpacs,nofootinbib,amsmath,amssymb]{revtex4}
\usepackage{amssymb,epsfig}
\usepackage[usenames,dvipsnames]{color}
\usepackage[bookmarks]{hyperref}

\newtheorem{lemma}{Lemma}
\newtheorem{proposition}{Proposition}
\newtheorem{theorem}{Theorem}

\newcommand{\proof}{\noindent {\bf Proof. }}

\newcommand{\qed}{\hfill $\fbox{\hspace{0.3mm}}$ \vspace{.3cm}} 

\newcommand{\Real}{\mathbb{R}}

\begin{document}

\title{Conformal diagrams for the gravitational collapse of a spherical dust cloud}
\author{N\'estor Ortiz and Olivier Sarbach}
\affiliation{Instituto de F\'\i sica y Matem\'aticas,
Universidad Michoacana de San Nicol\'as de Hidalgo,\\
Edificio C-3, Ciudad Universitaria, 58040 Morelia, Michoac\'an, M\'exico.}

\begin{abstract}
We present an algorithm for the construction of conformal coordinates in the interior of a spherically symmetric, collapsing matter cloud in general relativity. This algorithm is based on the numerical integration of the radial null geodesics and a local analysis of their behavior close to the singularity. As an application, we consider a collapsing spherical dust cloud, generate the corresponding conformal diagram and analyze the structure of the resulting singularity. A new bound on the initial data which guarantees that the singularity is visible from future null infinity is also obtained.
\end{abstract}

\date{\today}


\maketitle

\section{Introduction}

Once a sufficiently large mass is concentrated in a small region of an asymptotically flat spacetime, as occurs in the complete gravitational collapse of a star, for instance, a trapped surface appears~\cite{rSsY83} and the singularity theorems predict that a spacetime singularity forms (see~\cite{HawkingEllis-Book,Wald-Book} and references therein). An interesting question is whether this singularity is naked, or if it is hidden inside a black hole, such that no information escaping from it -- light rays in particular -- can be detected at future null infinity.

This question is essentially the contents of the weak cosmic censorship conjecture~\cite{rP69} which states that under suitable assumptions on the matter fields, the maximal Cauchy development of asymptotically flat, nonsingular initial data generically yields an asymptotically flat spacetime with a complete future null infinity. In case a naked singularity forms, there is a Cauchy horizon, and if it extends all the way to future null infinity it is not possible to predict the evolution of a test field in the asymptotic region for arbitrarily large times. As a consequence, future null infinity is incomplete and if this situation persists for perturbations of the initial data, weak cosmic censorship is violated. So far, despite much work, no general proof or disproof of this conjecture has been given. Reviews on this topic can be found in Refs.~\cite{rW97,pJ00}.

In this paper we analyze the gravitational collapse in the simple case of the Tolman-Bondi models, describing a collapsing spherically symmetric star with zero pressure. In these models spacetime is known in closed, explicit form, which greatly simplifies the analysis of its causal structure. In fact, it has been known for a long time that such collapse models lead to the formation of shell-focusing singularities, a portion of which is null and visible to local observers~\cite{pYhShM73,dElS79,dC84,rN86,pJiD93,Joshi-Book}. For appropriate initial data, part of the null singularity is even visible to observers which are arbitrarily far away from the dust cloud~\cite{dC84}, and there is a Cauchy horizon which extends all the way to future null infinity. We are interested in understanding how generic this feature is, at least within the class of Tolman-Bondi models.

To this purpose, we develop a method that generates a conformal diagram inside the collapsing, spherical dust cloud from given data for the initial density and velocity distributions. This method provides a valuable tool for understanding the causal structure of the spacetime. In particular, it enables one to determine in a systematic way whether or not a given initial data set results in a singularity that is hidden inside a black hole. Our method is based on a combination of analytic and numerical techniques. Analytic tools are used to understand the behavior of the null geodesics in the vicinity of the singularity, while numerical techniques are used to integrate the light rays away from the singularities. By generating the conformal diagrams for different initial data sets, we find that it is possible to obtain spacetimes with naked singularities which are globally visible without fine-tuning, indicating that these are generic within the class of spherically symmetric dust collapse. We also identify a large new class of initial data which leads to the formation of such globally naked singularities.

This paper is organized as follows. In section~\ref{Sec:Model}, we briefly review the Tolman-Bondi model in the bounded case, which describes the complete gravitational collapse of a spherical dust cloud, and state our assumptions on the initial data. In section~\ref{Sec:Theorems}, we start with a qualitative analysis of the in- and outgoing radial light rays emanating from or terminating in the shell-focusing singularity. In particular, we analyze the existence, uniqueness and asymptotic properties of such rays in a vicinity of the singularity. Our presentation is self-contained and presents a simple derivation of many known results. However, it also goes beyond previous results in the literature insofar that we obtain new asymptotic expansions for the light rays terminating at the singularity and a new bound on the initial data which guarantees that the resulting spacetime contains a globally visible singularity. Next, in section~\ref{Sec:Diagrams}, we describe our method for constructing the conformal coordinates inside the collapsing cloud. These coordinates provide a natural extension to the inside of the cloud of the Penrose-Kruskal coordinates for the Schwarzschild metric. Then, in section~\ref{Sec:Results}, we present the conformal diagrams corresponding to different initial data, and analyze in which cases the resulting singularity is naked or covered by an event horizon. Conclusions are drawn in section~\ref{Sec:Conclusions} and more technical points are discussed in the appendices.

\section{Tolman-Bondi dust collapse}
\label{Sec:Model}

In terms of co-moving, synchronous coordinates~\cite{MTW-Book}, the spacetime metric ${\bf g}$, four-velocity ${\bf u}$ and density $\rho$ for the solutions of the Einstein-Euler equations describing the gravitational collapse of a spherically symmetric dust cloud are given by
\begin{eqnarray}
{\bf g} &=& -d\tau^2 + \frac{r'(\tau,R)^2}{1 + 2E(R)}\; dR^2
 + r(\tau,R)^2(d\vartheta^2 + \sin^2\vartheta\, d\varphi^2),
\label{Eq:MetricSol}\\
{\bf u} &=& \frac{\partial}{\partial\tau}\; , \qquad
\rho(\tau,R) = \rho_0(R)\left( \frac{R}{r(\tau,R)} \right)^2\frac{1}{r'(\tau,R)}\; ,
\label{Eq:FluidSol}
\end{eqnarray}
where here the function $\tau\mapsto r(\tau,R)$ describes the evolution of the areal radius along the dust shell $R$ as a function of proper time, and $\dot{r}$ and $r'$ denote the partial derivatives of $r$ with respect to $\tau$ and $R$, respectively. We choose $R$ such that each dust shell is labeled by its initial areal radius at $\tau=0$, that is, $r(0,R) = R$. The time evolution of $r$ is governed by the one-dimensional mechanical system
\begin{equation}
\frac{1}{2} \dot{r}(\tau,R)^2 + V(r(\tau,R), R) = E(R),\qquad
V(r,R) := -\frac{m(R)}{r},
\label{Eq:1DMechanical}
\end{equation}
for each shell $R$, where $m(R)$ is the Misner-Sharp mass function~\cite{cMdS64} which is determined by the initial density profile $\rho_0$ according to
\begin{displaymath}
m(R) = 4\pi G\int\limits_0^R \rho_0(\bar{R})\bar{R}^2 d\bar{R},
\end{displaymath}
with Newton's constant $G$. The initial data consists of the initial velocity and density profiles $v_0(R) := \dot{r}(0,R)$ and $\rho_0(R)$, respectively, which fix the energy $E(R) = v_0(R)^2/2 - m(R)/R$ for each shell $R$.

We consider collapsing clouds of finite radius $R_1 > 0$. More precisely, our assumptions on the initial data are the following:
\begin{enumerate}
\item[(i)] $\rho_0,v_0: (0,\infty) \to \Real$ posses even and odd $C^\infty$-extensions, respectively, on the real axis $\Real$ (regular, smooth initial data),
\item[(ii)] $\rho_0(R) > 0$ for $0\leq R < R_1$ and $\rho_0(R)=0$ for $R\geq R_1$ (finite, positive density cloud),
\item[(iii)] $\rho_0'(R)\leq 0$ for all $R > 0$ (monotonically decreasing density),
\item[(iv)] $2m(R)/R < 1$ for all $R > 0$ (absence of trapped surfaces on the initial slice).
\end{enumerate}
Notice that condition (iv) automatically implies that $1 + 2E(R) > 0$  for all $R\geq 0$, such that equation~(\ref{Eq:MetricSol}) does not exhibit any coordinate singularities as long as $r > 0$ and $r' > 0$. Next, we impose conditions on the initial velocity profile:
\begin{enumerate}
\item[(v)] $v_0(R)/R < 0$  for all $R\geq 0$ (collapsing cloud),
\item[(vi)] $(v_0(R)/R)^2 < 2m(R)/R^3$  for all $R\geq 0$ (bounded collapse).
\end{enumerate}
The condition (vi) means that initially, the potential energy dominates the kinetic one such that the total energy is negative, $E(R)/R^2 < 0$ for all $R\geq 0$. One could also consider initial data for which the initial velocity is zero at some points, on an interval or everywhere, in which case the data is time-symmetric. In this respect condition (v) does not represent a genuine restriction since after an arbitrarily small time such data will evolve into a configuration where (v) and (vi) are both satisfied. Nevertheless, our results also apply to the time-symmetric case by substituting $v_0=0$ in our formulae below. Notice that the Oppenheimer-Snyder collapse~\cite{jOhS39}, for which $E=0$ and the density is homogeneous, is not covered by our assumptions. However, this case could be recovered by approximation from data satisfying our conditions.

For the following, it is convenient to introduce the functions
\begin{displaymath}
c(R) := \frac{2m(R)}{R^3},\qquad
q(R) := \sqrt{E(R)/V(R,R)} = \sqrt{1 - \frac{R v_0(R)^2}{2m(R)}}.
\end{displaymath}
The first quantity is proportional to the the mean density within the dust shell $R$ while $q(R)^2$ is the ratio between the total and initial potential energy. According to assumption (i), these functions have even $C^\infty$-extensions on the real axis, and assumptions (ii) and (iii) guarantee that $c(R) > 0$, $c'(R)\leq 0$ while assumptions (v) and (vi) imply that $0 < q(R) < 1$ for all $R\geq 0$. In fact, our conclusions hold equally well if we replace assumption (iii) by the weaker condition:
\begin{enumerate}
\item[(iii)'] $c'(R)\leq 0$ for all $R > 0$ (monotonically decreasing mean density).
\end{enumerate}
Finally, we impose the following two restrictions on the function $q$. First,
\begin{enumerate}
\item[(vii)] $q'(R)\geq 0$ for all $R > 0$ (exclusion of shell-crossing singularities),
\end{enumerate}
which, together with condition (iii)', implies that $r'(R) > 0$ for all $R > 0$ and guarantees that no shell-crossing singularities form, see Ref.~\cite{rN86} and the remark below the proof of Lemma~\ref{Lem:Elementary} in the next section. In order to formulate the second condition on $q$, we first note that the functions $c'/R$ and $q'/R$ are bounded near $R=0$ and have even $C^\infty$-extensions on the real axis. Then, the second condition is
\begin{enumerate}
\item[(viii)] For all $R\geq 0$, we have $q'(R)/R > 0$ whenever $c'(R)/R = 0$ (non-degeneracy condition).
\end{enumerate}
In particular, this means that the central values of $q''$ and $c''$ cannot be both zero. As we will see in the next section, this condition implies the existence of light rays escaping from the central singularity and making it visible, at least to local observers. On the other hand, if the central values of $q''$ and $c''$ are both zero, one can show~\cite{rN86} that no such light rays exist. Therefore, the condition (viii) for $R=0$ is the key property that determines whether or not the central singularity is locally visible.

Under the assumptions (i)--(viii), the solution of equation~(\ref{Eq:1DMechanical}) is given by the explicit formula
\begin{equation}
r(\tau,R) 
 = \frac{R}{q(R)^2} \left[f^{-1}\left( f(q(R)) + \sqrt{c(R)}q(R)^3\tau \right) \right]^2,
\label{Eq:Sol}
\end{equation}
with the strictly decreasing function
\begin{displaymath}
f: [0,1] \to [0,\pi/2],\quad x\mapsto x\sqrt{1 - x^2} + \arccos(x),
\end{displaymath}
whose derivative is $f'(x) = -2x^2/\sqrt{1 - x^2}$, $0\leq x < 1$. $f$ is a $C^\infty$-function on the interval $[0,1)$. The solution is regular on the domain $R\geq 0$ and $0\leq\tau < \tau_s(R)$, where the boundary $\tau = \tau_s(R)$ describes the shell-focusing singularity which is defined by the vanishing of $r/R$. From equation~(\ref{Eq:Sol}), we obtain
\begin{equation}
\tau_s(R) = \frac{\frac{\pi}{2} - f(q(R))}{\sqrt{c(R)}q(R)^3},\qquad R\geq 0.
\label{Eq:taus}
\end{equation}
Since the density $\rho$ of the dust diverges, the Einstein field equations imply that the Ricci scalar diverges at the shell-focusing singularity, and therefore, the boundary points $\tau = \tau_s(R)$ represent a curvature singularity. The tidal forces are much stronger near such points than in the case of shell-crossing singularities~\cite{pSaL99}. Outside the cloud, $R > R_1$, the spacetime is isometric to a subset of the Schwarzschild-Kruskal manifold according to Birkhoff's theorem, see for example Ref.~\cite{Straumann-Book}.

It is worth noticing that the key equation~(\ref{Eq:1DMechanical}), describing the dynamics of the dust shells, is identical to its Newtonian counterpart if $\tau$ is identified with Newtonian (absolute) time. What makes the relativistic part much more interesting, however, is the analysis of the resulting causal structure of spacetime. A particular interesting question is whether or not there exist light rays emanating from the shell-focusing singularity which are able to escape to future null infinity. This is discussed next.

\section{Light rays emanating from the singularity: qualitative analysis}
\label{Sec:Theorems}

In this section, we analyze the behavior of the radial null geodesics in the vicinity of the shell-focusing singularity. While most of the results derived here are known in the literature, see for instance, Refs.~\cite{dC84,rN86,pJiD93}, and have been generalized to non-radial causal geodesics~\cite{sDpJiD02}, see also~\cite{tSpJ96,bNfM01} for the marginally bound case, our derivation offers an alternative, self-contained and simple presentation of the theory. In addition, we obtain new asymptotic expansions for the light rays which will be important for generating the conformal diagram in the next section, and new results concerning the global behavior of the Cauchy horizon.

We divide the singular points into the central singularity, $\Sigma_0 := \{ (\tau_s(0),0) \}$, and the remaining part $\Sigma:= \{ (\tau_s(R),R) :  0 < R \leq R_1 \}$. We first establish for each point $p\in \Sigma$ the existence of a unique pair of in- and outgoing radial light rays terminating at $p$. Next, we prove that under our assumptions there is a unique ingoing light ray terminating at $\Sigma_0$, whereas there are infinitely many outgoing light rays emanating from the central singularity $\Sigma_0$. Then, we discuss the global behavior of the Cauchy horizon and determine fairly general conditions on the initial data which guarantee that it lies outside the black hole region. In particular, we provide a new upper bound for the central density which implies that in the spacetimes developing from initial data satisfying this bound, the central singularity  is visible from future null infinity.

\begin{figure}[h!]
\begin{center}
\includegraphics[width=8cm]{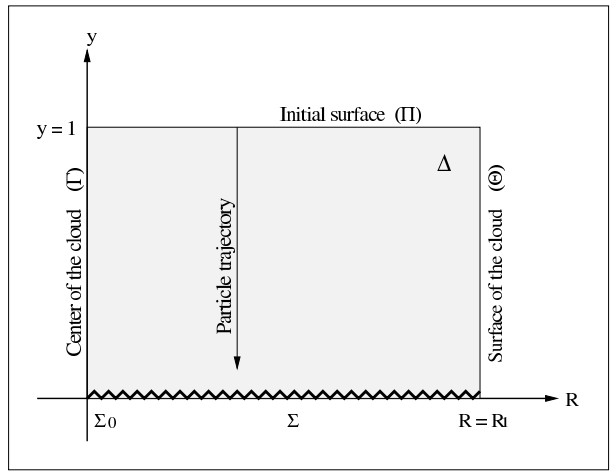}
\end{center}
\caption{\label{Fig:Rectangular_domain} A spacetime diagram representing the interior of the collapsing dust cloud in the coordinates $(y,R)$. In these coordinates, the interior is the rectangle $\Delta$ which is bounded by the initial surface $\Pi$, the surface of the cloud $\Theta$, the singularity $\Sigma\cup\Sigma_0$ and the center $\Gamma$.}
\end{figure}

Our method for analyzing the light rays is based on new local coordinates $(y,R)$, where $y$ is defined by $r(\tau,R) = R y^2$, that is, for each point $(\tau,R)$, $y^2$ is the ratio between the areal radii of the dust shell $R$ measured at times $\tau$ and $0$. With this, the spacetime domain inside the collapsing dust cloud we are interested in is the rectangular region
\begin{displaymath}
\Delta := \{ (y,R) : 0 < y < 1, 0 < R < R_1 \},
\end{displaymath}
whose boundary consists of the union of $\Sigma\cup \Sigma_0 = \{ (0,R) : 0\leq R \leq R_1 \}$ (the shell-focusing singularity), $\Theta:= \{ (y,R_1) : 0 < y \leq 1 \}$ (the surface of the cloud), $\Gamma := \{ (y,0) : 0 < y \leq 1 \}$ (the center of the cloud), and $\Pi := \{(1,R): 0\leq R \leq R_1 \}$ (the initial surface), see figure~\ref{Fig:Rectangular_domain}.

The radial null geodesics are the null curves of the radial part of the metric~(\ref{Eq:MetricSol}),
\begin{equation}
\frac{d\tau}{dR} = \epsilon \frac{r'(\tau,R)}{\sqrt{1 + 2E(R)}},
\label{Eq:dtau/dR}
\end{equation}
where $\epsilon = 1$ for outgoing and $\epsilon = -1$ for ingoing null geodesics. Expressed in terms of the new coordinates $(y,R)$, we have
\begin{equation}
\tau = \frac{g(q(R),y)}{\sqrt{c(R)}},\qquad
r' = y^2 + \frac{R^2}{y}\sqrt{1 - q(R)^2 y^2}\Lambda(y,R),
\label{Eq:taurprime}
\end{equation}
and the equation for the radial light rays is
\begin{equation}
\frac{dy}{dR} = \frac{1}{2}\sqrt{1-q(R)^2y^2}\left[ \frac{R\Lambda(y,R)}{y^2}
\left( 1 - \frac{\epsilon R Q(R)}{ y}\sqrt{1 - q(R)^2y^2} \right) - \epsilon Q(R) \right]
\label{Eq:dy/dR},
\end{equation}
where the functions $\Lambda: [0,1)\times [0,R_1] \to \Real$, $Q: [0,R_1]\to \Real$, and $g,h: (0,1)\times [0,1)\to\Real$ are defined as
\begin{eqnarray*}
\Lambda(y,R) &:=& 2\frac{q'(R)}{Rq(R)} h(q(R),y) 
  - \frac{c'(R)}{2Rc(R)} g(q(R),y),\\
Q(R) &:=& \sqrt{\frac{c(R)}{1 - R^2 q(R)^2 c(R)}},\\
g(q,y) &:=& \frac{ f(qy) - f(q) }{q^3},\\
h(q,y) &:=& \frac{1}{\sqrt{1-q^2}} - \frac{y^3}{\sqrt{1- q^2y^2}} - \frac{3}{2}g(q,y).
\end{eqnarray*}
Notice that $1 - R^2 q^2(R) c(R) = 1 + 2E(R) > 0$ and $c(R) > 0$ are positive, such that $Q$ is well-defined. In the time-symmetric case the first term in the expression for $\Lambda$ must be dropped, $g(q,y)$ is replaced by the function $f(y)$, and the function $h$ which is ill-defined for $q=1$ is not needed. It follows from our assumptions that the functions $\Lambda$, $Q$, $g$ and $h$ are $C^\infty$-differentiable. As a consequence, the coefficients on the right-hand side of equation~(\ref{Eq:dy/dR}) are smooth for all points $(y,R)\in\Delta$ in the interior of the cloud. However, equation~(\ref{Eq:dy/dR}) is singular at the shell-focusing singularity $y=0$ and the analysis of light rays emanating from or terminating at those points has to be treated specially.

Before undertaking this analysis, we summarize the elementary properties of the functions $\Lambda$, $Q$, $g$ and $h$ which will be important later:

\begin{lemma}
\label{Lem:Elementary}
The functions $\Lambda$, $Q$, $g$ and $h$ are strictly positive, $C^\infty$-differentiable functions on their domain. Furthermore, for fixed $q$, $g(q,\cdot)$ and $h(q,\cdot)$ are strictly decreasing and for fixed $R$, $\Lambda(\cdot,R)$ is strictly decreasing.
\end{lemma}

\proof
First, $Q > 0$ follows directly from its definition and our assumptions. Next, the positivity and monotonicity in $y$ of $g$ follow from the fact that $f$ is strictly decreasing. For the function $h$, we fix $0 < q < 1$ and note that
\begin{displaymath}
\lim\limits_{y\to 1} h(q,y) = 0,\qquad
\frac{\partial h}{\partial y}(q,y) = -\frac{q^2 y^4}{(1 - q^2 y^2)^{3/2}} < 0,\quad
0 < y < 1,
\end{displaymath}
which implies that $h(q,\cdot)$ is strictly decreasing and that $h(q,\cdot) > 0$ on the interval $[0,1)$. This, together with the assumptions (iii)', (vii) and (viii) immediately implies that $\Lambda > 0$ and that $\Lambda(\cdot,R)$ is strictly decreasing for fixed $R$.
\qed

In particular, it follows from equation~(\ref{Eq:taurprime}) and $\Lambda > 0$ that $r' > 0$ on $\Delta$, which precludes the existence of shell-crossing singularities, and implies that through each point $p\in\Delta$ there passes exactly one pair of in- and outgoing radial null geodesics.

In the following, we analyze the radial light rays through the singular points $p\in \Sigma$ and $p\in \Sigma_0$.

\subsection{Asymptotic behavior of the radial null rays near $\Sigma$}

Let $p = (0,R_0)\in \Sigma$ be a point on the non-central part of the singularity, $0 < R_0\leq R_1$. We show that there is a unique pair of in- and outgoing radial null rays terminating at $p$, which can be parametrized in the form $R = R_0 + \phi(y)$ for some $C^\infty$-function $\phi: [0,\delta)\to\Real$ satisying $\phi(0) = 0$. For this, we first rewrite equation~(\ref{Eq:dy/dR}) in its inverse form,
\begin{equation}
\frac{d\phi}{dy} = \frac{dR}{dy} = \frac{2}{\sqrt{1-q^2y^2}}\left[ \frac{R\Lambda}{y^2} \left( 1 - \frac{\epsilon R Q}{ y}\sqrt{1 - q^2y^2} \right) - \epsilon Q \right]^{-1}.
\label{Eq:dR/dy}
\end{equation}
In order to find the asymptotic behavior of the function $\phi$, we assume $\phi \approx A y^\alpha$ with $A\neq 0$ and $\alpha > 0$ some positive exponent to be determined. More precisely, following~\cite{pJiD93}, we set $x:=y^\alpha$ and assume $\phi(y) = \psi(x)$, with $\psi: [0,\delta)\to\Real$ a $C^1$-function satisfying $\psi(0)=0$ and $d\psi/dx(0) = A\neq 0$. The equation for the function $\psi$ implied by equation~(\ref{Eq:dR/dy}) is
\begin{displaymath}
\frac{d\psi}{dx} 
 = \frac{2}{\alpha}\frac{ x^{\frac{4}{\alpha}-1} }{\sqrt{1 - q^2 x^{\frac{2}{\alpha}} }}\left.
\left[ R\Lambda\left( y - \epsilon R Q\sqrt{1 - q^2 y^2} \right) - \epsilon Q y^3 \right]^{-1}
\right|_{y=x^{\frac{1}{\alpha}}, R = R_0 + \psi(x)}.
\end{displaymath}
Since the left-hand side converges to $A\neq 0$ while the expression inside the square parenthesis on the right-hand side converges to $-\epsilon R_0^2 Q(R_0)\Lambda(0,R_0)\neq 0$ for $x\to 0$, a necessary condition for the existence of such a solution is $\alpha=4$. In this case, we obtain, in the limit $x\to 0$,
\begin{equation}
A = -\frac{1}{2\epsilon R_0^2 Q(R_0)\Lambda(0,R_0)}.
\label{Eq:AValue}
\end{equation}

In order to give sufficient conditions for the existence of a solution, we make the ansatz
\begin{displaymath}
\phi(y) = A y^4 [1 + z(y)],
\end{displaymath}
where $A$ is given by equation~(\ref{Eq:AValue}) and $z: [0,\delta) \to \Real$ is a $C^\infty$-function satisfying $z(0) = 0$. As a consequence of equation~(\ref{Eq:dR/dy}), the function $z$ satisfies the differential equation
\begin{equation}
y\frac{dz}{dy} + 4z = y F(y,z),
\label{Eq:dz/dy}
\end{equation}
with the nonlinear term defined as
\begin{displaymath}
y F(y,z) := \left. \frac{2}{A\sqrt{1 - q^2 y^2} \left[R \Lambda y - \epsilon R^2 Q\Lambda\sqrt{1 - q^2 y^2} - \epsilon Q y^3 \right]} \right|_{R = R_0 + A y^4(1+z)} - 4.
\end{displaymath}
According to the definition of the constant $A$ and the elementary properties of the functions $q$, $Q$ and $\Lambda$, it follows that $y F: [0,\delta)\times (-\delta_1,\delta_1)\to\Real$ is a well-defined, $C^\infty$-function provided that $\delta > 0$ and $\delta_1 > 0$ are small enough. Since $y F(0,z) = 0$ for all $|z| < \delta_1$, 
it also follows that the function $F: [0,\delta)\times (-\delta_1,\delta_1)\to\Real$ itself is $C^\infty$.

Equation~(\ref{Eq:dz/dy}) is an ordinary differential equation with a regular singular point at $y=0$, with the nonlinear forcing term $y F(y,z)$. It follows the existence of a unique $C^\infty$-solution $z: [0,\varepsilon) \to \Real$ satisfying $z(0) = 0$ and $dz/dy(0) = F(0,0)/5$, see Theorem~1 in~\cite{dC84} or Theorem~\ref{Thm:RegSingPert} in Appendix~\ref{App:Theorem}. We summarize the main result of this subsection in:

\begin{proposition}
\label{Propo:LightRaysS}
Let $p = (0,R_0)\in \Sigma$. In the vicinity of $p$ there exists a unique pair of $C^1$ radial light rays terminating at $p$. Furthermore, these light rays have the form
\begin{equation}
R(y) = R_0 + A y^4[ 1 + z(y) ],
\label{Eq:LightRaysSExp}
\end{equation}
where $A$ is given by equation~(\ref{Eq:AValue}), and $z: [0,\delta) \to \Real$ is a $C^\infty$-function satisfying $z(0)=0$.
\end{proposition}

\proof Existence follows from Theorem~\ref{Thm:RegSingPert}, as discussed above. As for uniqueness, suppose $(y(\lambda),R(\lambda))$ is a $C^1$ radial light ray terminating at $p$. We may choose the parameter $\lambda$ such that the point $p$ corresponds to $\lambda=0$. Then, according to equation~(\ref{Eq:dR/dy}), we must have
\begin{displaymath}
\lim\limits_{\lambda\to 0} \frac{dR}{d(y^4)}
 = \lim\limits_{\lambda\to 0} \frac{dR}{4 y^3 dy} = A \neq 0,
\end{displaymath}
which implies that we can choose $\lambda=y$ in a vicinity $[0,\delta)$ of $p$. Furthermore, the $C^1$-function $z: (0,\delta) \to \Real$, $z(y):= (R(y)-R_0)/(A y^4) - 1$ is bounded and satisfies equation~(\ref{Eq:dz/dy}) and $\lim\limits_{y\to 0} z(y)=0$ according to l'H\^opital's rule. Now uniqueness follows from Theorem~\ref{Thm:RegSingPert}. Finally, it follows from equation~(\ref{Eq:dtau/dR}) and the sign of $A$ that $\tau$ increases as $y$ decreases to zero, showing that the point $p$ is the endpoint of the light rays.
\qed

\subsection{Asymptotic behavior of the radial null rays near $\Sigma_0$}

Next, we turn our attention to the radial light rays emanating from or terminating at the central singularity, $p = (0,0)\in \Sigma_0$. As in the previous subsection, we first try to find the asymptotic behavior by assuming that we can write the solution to equation~(\ref{Eq:dy/dR}) in the form $y(R) = \varphi(u)$, where $\varphi: [0,\delta)\to [0,\delta_1)$ is a $C^1$-function of the dimensionless variable $u:=(R/R_1)^\alpha$ satisfying $\varphi(0)=0$ and $d\varphi/du(0) = \lambda > 0$, with $\alpha > 0$ to be determined.

The function $\varphi$ satisfies the differential equation
\begin{equation}
\frac{d\varphi}{du} = \frac{R_1}{2\alpha}\sqrt{1- q(R)^2 y^2} \left[ 
R_1\Lambda(y,R) u^{\frac{2}{\alpha}-3}\left( \frac{u}{y} \right)^2
\left( 1 - \epsilon R_1 Q(R) u^{\frac{1}{\alpha}-1}\frac{u}{y}\sqrt{1 - q(R)^2 y^2} \right)
  - \epsilon Q(R) u^{\frac{1}{\alpha}-1} 
 \right]_{y = \varphi(u), R=R_1 u^{\frac{1}{\alpha}}}.
\label{Eq:dphi/du}
\end{equation} 
The left-hand side converges to $\lambda > 0$ in the limit $u\to 0$. The right-hand side also converges to a positive value if $\alpha=2/3$, in which case equation~(\ref{Eq:dphi/du}) simplifies to
\begin{equation}
\frac{d\varphi}{du} = \frac{3R_1}{4}\sqrt{1- q(R)^2 y^2} \left[ 
R_1\Lambda(y,R) \left( \frac{u}{y} \right)^2
\left( 1 - \epsilon R_1 Q(R)\sqrt{u}\frac{u}{y}\sqrt{1 - q(R)^2 y^2} \right)
  - \epsilon Q(R)\sqrt{u} \right]_{y = \varphi(u), R=R_1 u^{\frac{3}{2}}}.
\label{Eq:dphi/du2/3}
\end{equation} 
In the limit $u\to 0$ one obtains
\begin{equation}
\lambda = \left( \frac{3}{4}R_1^2\Lambda_0 \right)^{\frac{1}{3}} > 0,
\label{Eq:lambda}
\end{equation}
with
\begin{equation}
\Lambda_0 := \Lambda(0,0) =  \frac{2q_0''}{q_0} h(q_0,0) 
 - \frac{c_0''}{2c_0} g(q_0,0) > 0,
\label{Eq:Lambda_zero}
\end{equation}
where we have defined $q_0 := q(0)$, $q_0'' := q''(0)$, $c_0 := c(0)$ and $c_0'' := c''(0)$.

In order to prove the existence of such light rays, we make the ansatz $\varphi(x) = \lambda x^2 v(x)$, where $x := (R/R_1)^{1/3}$ and $v: [0,\delta)\to [0,\delta_1)$ is a $C^\infty$-function satisfying $v(0)=1$. The equation for $v$ follows easily from equation~(\ref{Eq:dphi/du}) with $\alpha=1/3$ and can be cast into the form
\begin{displaymath}
x\frac{dv^3}{dx} + 6v^3 = \frac{9R_1}{2} \sqrt{1-q(R)^2 y^2}\left[ \frac{R_1\Lambda(y,R)}{\lambda^3} \left( 1 - \frac{\epsilon R_1 Q(R) x}{\lambda v}\sqrt{1 - q(R)^2 y^2} \right) - \frac{\epsilon Q(R) x v^2}{\lambda} \right]_{R=R_1 x^{\frac{1}{3}},y = \lambda x^2 v}.
\end{displaymath}
Finally, we define a new $C^\infty$-function $z: [0,\delta)\to (-\delta_2,\delta_2)$ through $v(x)^3 = 1 + z(x)$, such that $z(0) = 0$. In terms of this function we have
\begin{equation}
x\frac{dz}{dx} + 6z = x F(x,z),
\label{Eq:dz/dx}
\end{equation}
with the nonlinear term
\begin{eqnarray*}
&& x F(x,z) \\
&& := \frac{9R_1}{2} \sqrt{1-q(R)^2 y^2}\left[ \frac{R_1\Lambda(y,R)}{\lambda^3} \left( 1 - \frac{\epsilon R_1 Q(R) x}{\lambda(1+z)^{1/3}}\sqrt{1 - q(R)^2 y^2} \right) 
 - \frac{ \epsilon Q(R) x(1+z)^{2/3}}{\lambda} \right]_{R=R_1 x^{\frac{1}{3}},y = \lambda x^2(1 +z)^{1/3}} - 6.
\end{eqnarray*}
Since $x F(x,z)$ vanishes identically for $x=0$ according to the definition of $\lambda$ in equation~(\ref{Eq:lambda}), it follows that $F: [0,\delta) \times (-\delta_2,\delta_2)\to \Real$ defines a $C^\infty$-function, and we can make use of Theorem~\ref{Thm:RegSingPert} to conclude the existence of a unique, local solution $z: [0,\varepsilon)\to \Real$ of equation~(\ref{Eq:dz/dx}) such that $z(0)=0$ and $dz/dx(0) = F(0,0)/7$. Since by equation~(\ref{Eq:dtau/dR}) $\tau$ increases (decreases) with $R$ for outgoing (ingoing) radial null rays, this demonstrates the existence of an outgoing radial light ray emanating from $\Sigma_0$ and the existence of an ingoing radial light ray terminating at $\Sigma_0$. We summarize our findings in

\begin{proposition}
\label{Propo:LightRaysS0Existence}
Let $p = (0,0)\in \Sigma_0$ be the central singularity. There exists an outgoing radial light ray emanating at $p$ and an ingoing radial light ray terminating at $p$, which have the form
\begin{equation}
y(x) = \lambda x^2 [ 1 + z(x) ]^{1/3},\qquad x = \left( \frac{R}{R_1} \right)^{1/3},
\label{Eq:LightRaysS0Exp}
\end{equation}
where the constant $\lambda$ is given by equation~(\ref{Eq:lambda}) and $z: [0,\delta)\to\Real$ is a $C^\infty$-function satisfying $z(0)=0$.

Furthermore, these rays are the unique $C^1$ radial light rays $y: [0,\delta)\to \Real$ satisfying $\lim\limits_{x\to 0} y(x)/x^2 = \lambda$.
\end{proposition}

The question of uniqueness is more subtle than in the previous subsection, and deserves a detailed discussion. The reason for this relies in the fact that there might exist radial light rays $y: [0,\delta) \to \Real$ with $y(0)=0$ whose limit $y(x)/x^2$ for $x\to 0$ does not exist. In order to analyze this, we start with the following technical lemma which is proven in Appendix~\ref{App:Lemma},

\begin{lemma}
\label{Lemma:uniqueness}
Let $\varphi: I := (0,\delta) \to (0,\delta_1)$ be a local, $C^1$-solution of equation~(\ref{Eq:dphi/du2/3}) such that $\lim\limits_{u\to 0} \varphi(u) = 0$. Define
\begin{displaymath}
m := \inf\limits_{u\in I} \frac{\varphi(u)}{u} \geq 0,\qquad
M := \sup\limits_{u\in I} \frac{\varphi(u)}{u} \leq \infty.
\end{displaymath}
Then, the following statements hold:
\begin{enumerate}
\item[(i)] If $m > 0$, then $M < \infty$.
\item[(ii)] If $M < \infty$ and $\epsilon = -1$, then $m > 0$.
\item[(iii)] If $m > 0$ and $M < \infty$, then $\lim\limits_{u\to 0} \frac{\varphi(u)}{u} = \lambda$, where $\lambda$ is defined by equation~(\ref{Eq:lambda}).
\end{enumerate}
\end{lemma}

As a consequence of this lemma, we have the following uniqueness result:

\begin{proposition}
\label{Propo:LightRaysS0Uniqueness}
Let $\varphi_0 : I := (0,\delta) \to (0,\delta_1)$ be the solution of equation~(\ref{Eq:dphi/du2/3}) of the form $\varphi_0(u) = \lambda u v_0(u)$ with $v_0: I \to \Real$ a $C^1$-function such that $\lim\limits_{u\to 0} v_0(u) = 1$, whose existence was proven in Proposition~\ref{Propo:LightRaysS0Existence}. Let $\varphi: I':=(0,\delta') \to (0,\delta'_1)$ be a local, $C^1$-solution of equation~(\ref{Eq:dphi/du2/3}) with $0 < \delta' \leq \delta$ and $0 < \delta'_1 \leq \delta_1 $ such that $\lim\limits_{u\to 0}\varphi(u) = 0$.

Then, in the ingoing case $\epsilon = -1$, it follows that $\varphi(u) = \varphi_0(u)$ for all $u\in I'$, that is, $\varphi_0$ is the unique local solution which connects $\Sigma_0$. In the outgoing case $\epsilon = 1$, it follows that $\varphi(u) \leq \varphi_0(u)$ for all $u\in I'$, that is, $\varphi_0$ is the earliest radial light ray escaping from $\Sigma_0$.
\end{proposition}

\proof Since $\lim\limits_{u\to 0} v_0(u) = 1$ there are constants $M > m > 0$ such that $m \leq \lambda v_0(u) \leq M$ for all $u\in I$. Let us define the  function $v(u) := \varphi(u)/(\lambda u)$, $u\in I'$. According to Proposition~\ref{Propo:LightRaysS0Existence}, it follows that $v(u) = v_0(u)$ for all $u\in I'$ if we can show that $\lim\limits_{u\to 0} v(u) = 1$.

Now consider the case $\epsilon = -1$ first, and suppose $\varphi(u)\neq \varphi_0(u)$ for some $u\in I'$. Since the solutions cannot cross on $\Delta$, it follows that either $\varphi(u) > \varphi_0(u)$ for all $u\in I'$ or $\varphi(u) < \varphi_0(u)$ for all $u\in I'$. In the first case, it follows that $\varphi(u)/u > \varphi_0(u)/u = \lambda v_0(u) \geq m > 0$ for all $u\in I'$. Then, Lemma~\ref{Lemma:uniqueness}(i) and (iii) imply that $\lim\limits_{u\to 0} \varphi(u)/u = \lambda$, and uniqueness follows. In the second case, $\varphi(u)/u < \varphi_0(u)/u = \lambda v_0(u) \leq M$ for all $u\in I'$ and Lemma~\ref{Lemma:uniqueness}(ii) and (iii) imply that $\lim\limits_{u\to 0} \varphi(u)/u = \lambda$.

Finally, suppose $\epsilon = 1$ and $\varphi(u) > \varphi_0(u)$. Then, Lemma~\ref{Lemma:uniqueness}(i) and (iii) imply that $\lim\limits_{u\to 0} \varphi(u)/u = \lambda$ as before, and uniqueness follows.
\qed

Notice that in the case of outgoing radial null geodesics emanating from $\Sigma_0$, we cannot prove that $\varphi_0(u)$ is unique. In fact, it turns out that there are infinitely many outgoing radial light rays emanating from $\Sigma_0$. Indeed, given a point $p = (R_0,y_0)\in\Delta$ sufficiently close to $\Sigma_0$ and such that $y_0 < \varphi_0(u_0)$, $u_0 = (R_0/R_1)^{2/3}$, the outgoing light ray $\varphi(u)$ passing through this point cannot cross $\varphi_0$, and so $\varphi(u)/u < \varphi_0(u)/u \leq M$ for all $u\in I$ for which $\varphi$ is defined. Since by Propostion~\ref{Propo:LightRaysS} the light ray $\varphi$ cannot emanate from $\Sigma$, it follows that $\varphi(u)$ connects $\Sigma_0$. Therefore, there are infinitely many radial light rays emanating from $\Sigma_0$, the earliest of which is $\varphi_0$, which generates a Cauchy horizon.

We summarize the results obtained so far in the following

\begin{theorem}
\label{Thm:CH}
Given the assumptions (i),(ii),(iii)',(iv)--(viii) made in section~\ref{Sec:Model}, we have the following behavior for the radial null geodesics near the shell-focusing singularity: for each point $p = (0,R_0)\in \Sigma$, $R_0 > 0$, there exists  a unique pair of $C^\infty$ radial light rays terminating at $p$. There is also a unique, $C^\infty$ incoming radial light ray terminating at the central singularity $(0,0)\in \Sigma_0$. However, there are infinitely many outgoing light rays emanating from $\Sigma_0$. The earliest of those is described by the solution given in Proposition~\ref{Propo:LightRaysS0Existence} and generates the Cauchy horizon.
\end{theorem}

These results, including the asymptotic expansions~(\ref{Eq:LightRaysSExp},\ref{Eq:LightRaysS0Exp}) turn out to be important ingredients for the algorithm generating the conformal diagrams described in section~\ref{Sec:Diagrams}. However, before discussing this algorithm, we analyze conditions for which the light ray generating the Cauchy horizon arrives at the surface of the dust cloud earlier than the event horizon, implying that part of the singularity is visible to outside observers.

\subsection{Global visibility of the singularity}

Here, we find conditions on the initial data which guarantee the formation of a naked singularity which is visible not only to local observers, but also to observers at an arbitrarily large distance from the cloud. This occurs if and only if the light ray generating the Cauchy horizon arrives at the surface of the cloud earlier than the event horizon, implying the existence of light rays emanating from the singularity which reach the Schwarzschild exterior spacetime at $r > 2m$.

In order to analyze this question, it is useful to consider the apparent horizon. In our spherically symmetric model, it is determined by the critical surface dividing the two regions in which the areal radius is increased and decreased, respectively, along the outgoing radial null rays. Parametrizing the outgoing radial null rays by $R$, and using equations~(\ref{Eq:dtau/dR}), (\ref{Eq:1DMechanical}) and (\ref{Eq:taurprime}) we obtain
\begin{equation}
\frac{dr}{dR} = \frac{d\tau}{dR}\dot{r} + r'
 = \left(y^2 + R^2\Lambda(y,R)\frac{\sqrt{1 - q(R)^2y^2}}{y}  \right)
  \left(1 - RQ(R)\frac{\sqrt{1 - q(R)^2y^2}}{y} \right)
\label{Eq:dr/dR}
\end{equation}
along the outgoing null rays, where $y^2 = r/R$, as before. It follows that the apparent horizon is the surface for which the expression inside the second parenthesis vanishes, which is equivalent to the condition $y = y_{AH}(R) := R\sqrt{c(R)}$ or
\begin{equation}
r = r_{AH}(R) := R^3 c(R) = 2m(R).
\end{equation}
From equation~(\ref{Eq:LightRaysS0Exp}), we see that at least for $R\ll R_1$ the light ray $y = y_0(R)$ generating the Cauchy horizon lies outside the apparent horizon, since $y_0(R) \simeq R^{2/3}$ while $y_{AH}(R)\simeq R$. Next, we also note that an outgoing null ray emanating from a point inside the apparent horizon, $r < 2m(R)$, cannot escape this region since $r$ decreases while $m(R)$ increases as $R$ grows. Therefore, a null ray emanating from a point inside the apparent horizon inside the cloud either reaches the singularity inside the cloud, or the surface of the cloud with $r < 2m(R_1)$, and hence it lies inside the black hole region. As a consequence, a necessary and sufficient condition for the Cauchy horizon to lie outside the event horizon is that the light ray $y_0$ lies {\em outside} the apparent horizon for all $R > 0$, that is,
\begin{displaymath}
r_0(R) := R y_0(R)^2 > 2m(R)
\end{displaymath}
for all $R > 0$. In the following, we establish that under suitable assumptions on the initial data this estimate holds. This is achieved by estimating each term in the propagation equation~(\ref{Eq:dr/dR}) for the areal radius. We start with the following simple observation.

\begin{lemma}
One has the following upper bound for the function $y_0$, describing the generator of the Cauchy horizon:
\begin{equation}
y_0(R) \leq \frac{1}{q(R)} f^{-1}\left( 
 f(q(R)) + \left[ \frac{\pi}{2} - f(q_0) \right]\frac{q(R)^3}{q_0^3}\sqrt{\tilde{c}(R)} 
 \right) =: \eta(R),
\end{equation}
for all $0 \leq R \leq R_1$, where $\tilde{c}(R) := c(R)/c_0$ is the normalized mean density profile and where $c_0 := c(0)$ and $q_0 := q(0)$.
\end{lemma}

\proof Since by equation~(\ref{Eq:dtau/dR}) $\tau$ cannot decrease along $y_0$, we have, using equation~(\ref{Eq:taurprime}),
\begin{displaymath}
g(q(R),y_0(R)) = \sqrt{c(R)} \tau \geq \sqrt{c(R)}\tau_s(0),
\end{displaymath}
where according to equation~(\ref{Eq:taus}), $\tau_s(0) = (\pi/2 - f(q_0))/(q_0^3\sqrt{c_0})$. Now the statement of the lemma follows from the definition of the function $g$ and the fact that $f$ is strictly monotonically decreasing.
\qed

Notice that $\eta(R)$ is strictly positive for $R > 0$ and satisfies $\eta(R)\simeq R^{2/3}$ for $R\ll R_1$. Furthermore, $\eta$ only depends on the initial function $q$ and the initial normalized mean density $\tilde{c} = c/c_0$, but not on its magnitude $c_0$.

Next, assuming that the areal radius is uniformly bounded away from the apparent horizon on some interval, we integrate equation~(\ref{Eq:dr/dR}) to obtain an appropriate lower bound on $r$:

\begin{lemma}
\label{Lem:CHEstimate}
Let $\delta > 0$, and suppose $y_0(R) \geq (1 + \delta) y_{AH}(R)$ for all $0\leq R \leq R_0$. Then, the generator of the Cauchy horizon satisfies
\begin{equation}
r_0(R) \geq \frac{\delta}{1 + \delta}\xi(R) 
 + \delta(1 + \delta)\int\limits_0^RÊ\bar{R}^2 c(\bar{R}) d\bar{R},
\qquad
\xi(R) := \int\limits_0^R \bar{R}^2\Lambda(\eta(\bar{R}),\bar{R})
\frac{\sqrt{1 - q(\bar{R})^2\eta(\bar{R})^2}}{\eta(\bar{R})} d\bar{R}
\end{equation}
for all $0\leq R \leq R_0$.
\end{lemma}

\proof
We estimate the expressions inside the two parenthesis on the right-hand side of equation~(\ref{Eq:dr/dR}) from below. For the first expression, we note that $y^2 \geq (1+\delta)^2 R^2 c(R)$, $\Lambda(y,R) \geq \Lambda(\eta(R),R)$ and $\sqrt{1 - q(R)^2 y^2}/y \geq \sqrt{1 - q(R)^2\eta(R)^2}/\eta(R)$, where we have used the results from the previous Lemma and Lemma~\ref{Lem:Elementary}. For the second expression we use
\begin{displaymath}
RQ(R)\frac{\sqrt{1 - q(R)^2y^2}}{y} 
 \leq R Q(R)\frac{\sqrt{1 - (1+\delta)^2 R^2 q(R)^2 c(R)}}{(1+\delta) R\sqrt{c(R)}}
 = \sqrt{\frac{1 - (1+\delta)^2 R^2 q(R)^2 c(R)}{1 - R^2 q(R)^2 c(R)}}\frac{1}{1 + \delta}
 \leq \frac{1}{1 + \delta}.
\end{displaymath}
Putting everything together, we obtain
\begin{displaymath}
\frac{dr}{dR} \geq \left[ (1+\delta)^2 R^2 c(R) 
 + R^2\Lambda(\eta(R),R)\frac{\sqrt{1 - q(R)^2\eta(R)^2}}{\eta(R)} \right]
 \frac{\delta}{1 + \delta}
\end{displaymath}
for all $0\leq R \leq R_0$. Integrating both sides of this inequality and observing that $r(0)=0$, the statement follows immediately.
\qed

Like $\eta(R)$, the function $\xi(R)$ is strictly positive for $R > 0$ and only depends on $q$ and the normalized mean density $\tilde{c} = c/c_0$. In contrast to this, the mass function $m(R) = R^3 c(R)/2$ scales with the magnitude $c_0$ of the central density. Therefore, under the hypothesis of Lemma~\ref{Lem:CHEstimate}, we can always arrange for $r_0(R) > 2m(R)$ on the interval $0\leq R\leq R_0$ by rescaling the initial density profile. This observation leads to our final result on the global behavior of the Cauchy horizon.

\begin{theorem}
\label{Thm:Naked}
Consider an initial density and velocity profile $(\tilde{\rho}_0,\tilde{v}_0)$ satisfying $\tilde{c}(0) = 1$ and the conditions (i),(ii),(iii)',(iv)--(viii) in section~\ref{Sec:Model}. Let $\mu > 0$ be sufficiently small such that $\mu\leq 1$ and
\begin{equation}
\mu^2 \leq \inf\limits_{0 < R \leq R_1} \frac{4}{27}\frac{\xi(R)}{R^3\tilde{c}(R)}.
\label{Eq:c0Bound}
\end{equation}
Then, the rescaled data $(\rho_0:=\mu^2\tilde{\rho}_0,v_0:=\mu\tilde{v}_0)$ also satisfies the conditions (i),(ii),(iii)',(iv)--(viii) in section~\ref{Sec:Model}, and the generator of the Cauchy horizon in the resulting spacetime satisfies
\begin{displaymath}
r_0(R_1) \geq \frac{9}{2} m(R_1) 
 + \frac{3\mu^2}{4}\int\limits_0^{R_1}Ê\bar{R}^2\tilde{c}(\bar{R}) d\bar{R} > 2m(R_1).
\end{displaymath}
\end{theorem}

\proof
First, we note that $\xi(R)\simeq R^{7/3}$ for $R\ll R_1$, and that $\xi(R)$ is strictly positive for $R > 0$, such that the function $\xi(R)/(R^3\tilde{c}(R))$ is strictly positive and diverges for $R\to 0$. Therefore, the infimum in the bound~(\ref{Eq:c0Bound}) is strictly positive.

Next, we note that the rescaled solution satisfies $c=\mu^2\tilde{c}$ and $q=\tilde{q}$, since the rescaling is such that both the initial kinetic and potential energies scale with $\mu^2$. Now fix\footnote{Other positive values for $\delta$ could also be considered. Our choice is the one that maximizes the constant in the bound~(\ref{Eq:c0Bound}).} $\delta = 1/2$ and let $R_0\in [0,R_1]$ be the maximum value for which $y_0(R)\geq (1 + \delta) y_{AH}(R)$ for all $0\leq R \leq R_0$. Observe that $R_0 > 0$ since $y_0(R)\simeq R^{2/3}$ whereas $y_{AH}(R)\simeq R$ for $R\ll R_1$. Now Lemma~\ref{Lem:CHEstimate} and the hypothesis imply that
\begin{displaymath}
r_0(R) \geq \frac{\delta}{1 + \delta}\xi(R) 
 + \delta(1 + \delta)\int\limits_0^RÊ\bar{R}^2 c(\bar{R}) d\bar{R}
 \geq \frac{9}{4} R^3 c(R) + \frac{3}{4}\int\limits_0^RÊ\bar{R}^2 c(\bar{R}) d\bar{R}
\end{displaymath}
for all $0\leq R \leq R_0$. In particular, this implies that $y_0(R_0) > 3 y_{AH}(R_0)/2$. By the maximality of $R_0$, this means that $R_0 = R_1$, and the proposition follows.
\qed

Therefore, given a normalized mean density profile $\tilde{c}$ and an initial ratio $q^2$ between total and potential energy satisfying our assumptions, the existence of a singularity which is locally visible is based on the requirement that the second derivatives $\tilde{c}''$ and $q''$ at the center $R=0$ do not vanish both (see the comments below assumption (viii) in section~\ref{Sec:Model}), while the global visibility is guaranteed if the magnitude $c_0$ of the mean density profile $c = c_0\tilde{c}$ is sufficiently small. For the case of time-symmetric initial data, where $q=1$, the expression for the function $\xi$ simplifies to
\begin{displaymath}
\xi(R) = \int\limits_0^R \bar{R}\frac{d}{d\bar{R}}\eta(\bar{R})^2 d\bar{R}
 = R\eta(R)^2 - \int\limits_0^R \eta(\bar{R})^2 d\bar{R},\qquad
\eta(R) = f^{-1}\left( \frac{\pi}{2}\sqrt{\tilde{c}(R)} \right).
\end{displaymath}

In the next section, we numerically compute the bound~(\ref{Eq:c0Bound}) for a four-parameter family of initial data and show that it is consistent with the results obtained from the conformal diagrams.

\section{Light rays emanating from the singularity: quantitative study and conformal diagrams}
\label{Sec:Diagrams}

In this section we present our algorithm for constructing conformal coordinates $(T,X)$, in which the radial part of the metric~(\ref{Eq:MetricSol}),
\begin{displaymath}
\tilde{\bf g} := -d\tau^2 + \frac{dR^2}{\gamma(\tau,R)^2},\qquad
\gamma(\tau,R) := \frac{\sqrt{1 + 2E(R)}}{r'(\tau,R)},
\end{displaymath}
assumes the simple form
\begin{displaymath}
\tilde{\bf g} = \Omega(T,X)^2( -dT^2 + dX^2),
\end{displaymath}
with $\Omega(T,X) > 0$ the conformal factor. In these coordinates, the causal structure is transparent since the radial light rays are simply described by the straight lines $T\pm X = const.$ The coordinates $T$ and $X$ are conveniently obtained by introducing the null coordinates $U := T - X$ and $V := T + X$ which satisfy the advection equations
\begin{equation}
\dot{U} = -\gamma U',\qquad \dot{V} = +\gamma V',
\label{Eq:Advection}
\end{equation}
subject to appropriate boundary conditions. Indeed, if $U$ and $V$ satisfy equation~(\ref{Eq:Advection}), then
\begin{displaymath}
\Omega^2( -dT^2 + dX^2) = -\Omega^2 dU dV
= \Omega^2\dot{U}\dot{V}\left( -d\tau^2 + \frac{dR^2}{\gamma(\tau,R)^2} \right),
\end{displaymath}
and the conformal factor is $\Omega = 1/\sqrt{\dot{U}\dot{V}}$, provided that $\dot{U}$ and $\dot{V}$ are positive.

\subsection{The special case of a homogeneous cloud with zero initial velocity}

\begin{figure}[h!]
\begin{center}
\includegraphics[width=12cm]{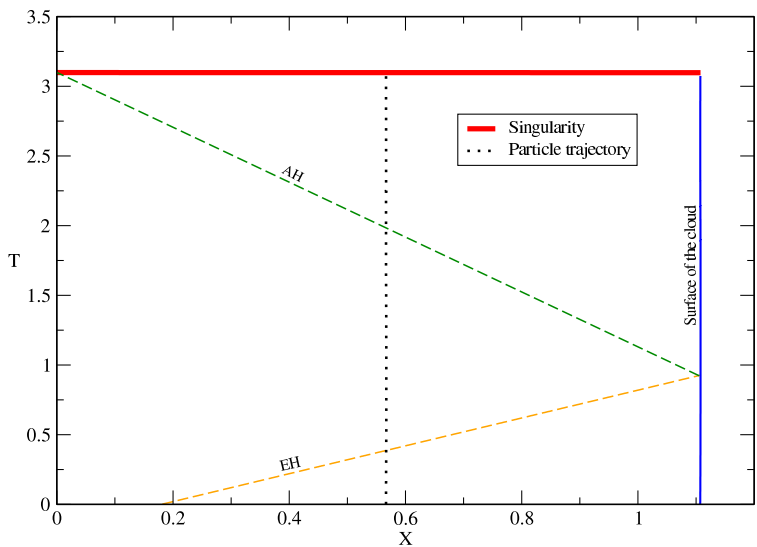}
\end{center}
\caption{\label{Fig:Homogeneous} Conformal diagram for a homogenous dust cloud with zero initial velocity and the parameter choices $c_0 = 0.75$ and $R_1=1$. The lines denoted by ``AH" and ``EH" refer to the apparent and event horizons, respectively. Since outside the cloud the apparent horizon coincides with the event horizon, the event horizon inside the cloud is determined by the outgoing null ray passing through the intersection of the apparent horizon with the surface of the cloud. Notice also that for these parameter values, the event horizon intersects the initial surface, implying that the dust particles which are close to the center initially, are already imprisoned inside the black hole region.}
\end{figure}

As a simple example, consider first the case of homogeneous density and zero initial velocity, for which $q=1$ and $c = c_0 = const$. In this case, $r = R y^2$, $\gamma^{-1} = y^2/\sqrt{1 - c_0 R^2}$, where $y = f^{-1}(\sqrt{c_0}\tau)$ is independent of $R$. Therefore,
\begin{displaymath}
\tilde{\bf g} = \frac{y^4}{c_0}\left[ -\left( \frac{2dy}{\sqrt{1 - y^2}} \right)^2 
 + \left( \frac{\sqrt{c_0} dR}{\sqrt{1 - c_0 R^2}} \right)^2 \right],
\end{displaymath}
and the substitutions $y = \cos(T/2)$, $\sqrt{c_0} R = \sin(X)$ lead to the Friedman-Robertson-Walker form of the metric with conformal time $T$,
\begin{displaymath}
{\bf g} = \frac{\cos^4(T/2)}{c_0}\left[ -dT^2 + dX^2 
 + \sin^2(X)(d\vartheta^2 + \sin^2\vartheta\, d\varphi^2)\right],\qquad
0\leq T \leq \pi,\quad 0\leq X \leq \arcsin(\sqrt{c_0} R_1).
\end{displaymath}
The corresponding conformal diagram for $c_0 = 0.75$ and $R_1 = 1$ is shown in figure~\ref{Fig:Homogeneous}. Notice that the apparent horizon equation $y_{AH}(R) = \sqrt{c_0} R$ reduces to the simple equation $T = \pi - 2X$, which describes a time-like three-surface. The singularity is spacelike and hidden inside the black hole region. Surprisingly, the picture changes completely in the generic case. As is already clear from Theorem~\ref{Thm:CH}, the central singularity is null and visible to local observers when the assumptions (i)-(viii) are met. Furthermore, the null piece of the singularity may extend far enough into the past such that the Cauchy horizon lies outside the black hole, see Theorem~\ref{Thm:Naked} and the conformal diagrams below.

\subsection{The generic case}

In the generic case, where the initial velocity and density profiles satisfy the assumptions (i)--(viii), we solve the advection equations~(\ref{Eq:Advection}) by the method of characteristics. The equations~(\ref{Eq:Advection}) imply that $U$ and $V$ are constant along the out- ($\tau_+(R)$) and ingoing ($\tau_-(R)$) radial null geodesics, respectively, where
\begin{equation}
\frac{d\tau_+}{dR} = +\frac{1}{\gamma(\tau_+,R)},\qquad
\frac{d\tau_-}{dR} = -\frac{1}{\gamma(\tau_-,R)}.
\end{equation}
In terms of the coordinates $(y,R)$ introduced in section~\ref{Sec:Theorems}, these equations are
\begin{equation}
\frac{dy_+}{dR} = w_+(y_+,R),\qquad
\frac{dy_-}{dR} = w_-(y_-,R),
\end{equation}
where the function $w_\epsilon(y,R)$, $\epsilon = \pm 1$, is given by the right-hand side of equation~(\ref{Eq:dy/dR}). For the numerical integration we use, instead of $R$, a new parameter $\lambda$ and rewrite the equations for the radial null rays as the autonomous systems,
\begin{equation}
\frac{dy_\pm}{d\lambda} 
 = \frac{\pm w_\pm(y_\pm,R_\pm)}{\sqrt{1 + w_\pm(y_\pm,R_\pm)^2}}, \qquad
\frac{dR_\pm}{d\lambda} =  \frac{\pm 1}{\sqrt{1 + w_\pm(y_\pm,R_\pm)^2}}.
\label{Eq:ODE}
\end{equation}
The definition of the parameter $\lambda$ is such that it corresponds to the arc length with respect to the (artificial) Euclidean metric in the $(y,R)$-chart. Therefore, the right-hand sides of equations~(\ref{Eq:ODE}) cannot be very large or both very small, and there is no need for adaptive or implicit methods when numerically integrating the equations. As a consequence, it is sufficient to use a fourth-order Runge-Kutta time integrator (see, for instance, Ref.~\cite{Recipies-Book}) with fixed step size $h$ when integrating the ODE system~(\ref{Eq:ODE}). The sign of $\lambda$ is such that $\tau$ increases with $\lambda$ along the in- and outgoing radial light rays.

Once the characteristics are found, the null coordinates $U$ and $V$ can be constructed by assigning to each outgoing null ray a unique value for $U$ and to each ingoing null ray a unique value for $V$, such that $\dot{U} > 0$, $\dot{V} > 0$. For this, we specify boundary conditions for $U$ and $V$ at the surface $\Theta$ of the cloud, boundary conditions for $U$ at the singularity $\Sigma$ and boundary conditions for $V$ at the initial surface $\Pi$ (see figure~\ref{Fig:Rectangular_domain}). In order to assure that $\dot{U},\dot{V} > 0$ we require that $U$ increases as one moves along $\Theta$ to the future, and then along the singularity $\Sigma$ toward the center. Similarly, we ask that $V$ increases when one moves along $\Pi$ from the center to the surface, and then along $\Theta$ to the future.

In what follows, we describe these boundary conditions in more detail. The idea is to match the conformal diagram smoothly to the Penrose-Kruskal diagram of the exterior Schwarzschild solution. For a different choice of boundary conditions, see Ref.~\cite{nOoS10}.

\subsubsection{Boundary conditions for $U$ and $V$ at $\Theta$}

The quantities $U$ and $V$ at $\Theta$ are determined by matching their values to those of the Penrose-Kruskal coordinates in the Schwarzschild spacetime. For this, we obtain $U$ and $V$ analytically by computing the trajectory of a free falling, radial observer in the Schwarzschild spacetime which starts at $r = R_1$ with initial velocity $v_0(R_1)$. The calculation which is performed in Appendix~\ref{App:Kruskal} yields the following result:
\begin{eqnarray*}
\tan(U_{\Theta}(y)) &=& \frac{1}{a_1} \left( \sqrt{a_1^2-b_1^2y^2} - y\sqrt{1-b_1^2} \right) \\
 &\times& \exp \left\{ \frac{y^2}{2a_1^2} - \frac{\sqrt{1-b_1^2}}{2b_1^2}
\left[ \frac{1+2b_1^2}{b_1}\left( \arctan \left( \frac{\sqrt{a_1^2-b_1^2y^2}}{b_1y} \right)
 -\frac{\pi}{2} \right) +  \frac{y}{a_1^2}\sqrt{a_1^2-b_1^2y^2}  \right]\right\}, \\
\tan(V_{\Theta}(y)) &=& \frac{1}{a_1} \left( \sqrt{a_1^2-b_1^2y^2} + y\sqrt{1-b_1^2} \right)\\
 &\times& \exp \left\{ \frac{y^2}{2a_1^2} + \frac{\sqrt{1-b_1^2}}{2b_1^2}
\left[ \frac{1+2b_1^2}{b_1}\left( \arctan \left( \frac{\sqrt{a_1^2-b_1^2y^2}}{b_1y} \right) 
 - \frac{\pi}{2} \right) +  \frac{y}{a_1^2}\sqrt{a_1^2-b_1^2y^2}  \right] \right\},
\end{eqnarray*}
where $a_1 := \sqrt{2m(R_1)/R_1} = R_1\sqrt{c(R_1)}$ and $b_1 := \sqrt{-2E(R_1)} = q(R_1) a_1$. As shown in Appendix~\ref{App:Kruskal}, $U_\Theta$ and $V_\Theta$ are strictly decreasing functions of $y$ which are normalized such that $U_\Theta(0) = V_\Theta(0) = \arctan(1) = \pi/4$.

\subsubsection{Boundary conditions for $V$ at $\Pi$}

At the initial surface $\Pi$, we choose $V$ to be a linear function such that the values of it and its derivative match the corresponding expressions for the Penrose-Kruskal coordinates. Therefore,
\begin{displaymath}
V_\Pi(R) := V_\Theta(1) + \beta \left( \frac{R}{R_1} - 1 \right),\qquad
\beta := R_1 \left. \frac{\partial V_\Theta}{\partial R_1} \right|_{y=1}.
\end{displaymath}
For the cases studied below we have verified that $\beta$ is positive, implying that $V' > 0$ along $\Pi$.

\subsubsection{Boundary conditions for $U$ at $\Sigma$}

Our choice for $U$ at the spacelike portion $\Sigma$ of the singularity is based on the observation that in the Penrose diagram for the Schwarzschild spacetime, the curvature singularity at $r=0$ is a subset of the straight line $2T = U + V = const.$ Therefore, we choose $U$ such that $U + V = const. = U_\Theta(0) + V_\Theta(0) = \pi/2$. However, this requires the knowledge of $V$ at each point of $\Sigma$. We determine the value $V_p$ of $V$ at a given point $p\in \Sigma$ by computing numerically an ingoing radial null geodesics to the past until it intersects either the surface of the cloud $\Theta$ or the initial surface $\Pi$, where $V$ is known. Since $V$ is constant along the ingoing null ray, this determines $V$ at $p$. Then, $U$ is obtained by setting $U_p := \pi/2 - V_p$.

\subsection{Construction of the conformal diagram}

The boundary conditions determine a unique pair of coordinates $(U,V)$ for each point $p\in\Delta$ in the inside of the cloud. To find it, it is in principle sufficient to compute a future directed outgoing radial null ray and a past directed ingoing radial null ray emanating from $p$. The outgoing ray intersects either $\Sigma$ or $\Theta$, where $U$ is given. The ingoing ray intersects either $\Pi$ or $\Theta$, where $V$ is known. With this algorithm it is in principle possible to construct the images of the relevant surfaces $\Pi$, $\Gamma$, $\Theta$, $\Sigma$, the apparent horizon or a dust particle trajectory in the conformal $TX$ diagram, where $T = (V+U)/2$, $X = (V-U)/2$, by computing the coordinates $(U,V)$ for each point on these surfaces.

However, in practice, a problem with this procedure occurs when the outgoing rays 
approach the singular surface $\Sigma$ since at $\Sigma$ the equation for the null ray is undefined. In order to avoid integrating toward the singularity, we adopt the following algorithm:

\begin{enumerate}
\item[{\bf Step 1}.] We choose a uniform grid $(y_j,R_1)$ along the surface $\Theta$ of the cloud, where $y_j = j/J$, $j=0,1,2,...,J$. Since $U$ and $V$ are given at $\Theta$, we immediately obtain the conformal image of $\Theta$ by computing $(U_\Theta(y_j),V_\Theta(y_j))$ at all gridpoints $j=0,1,2,...,J$.
\item[{\bf Step 2}.] For each $j=J,J-1,...$ we shoot from $(y_j,R_1)$ an outgoing radial light ray to the past. This light ray may intersect one of the surfaces $\Pi$, $\Gamma$, the apparent horizon or a dust particle trajectory at a point $p$, say. Then, we  associate to $p$ the value $U_p = U_\Theta(y_j)$ for $U$, and determine $V_p$ by shooting from $p$ an ingoing radial light ray directed to the past, as described above. This gives the conformal image of the relevant surfaces in the past of the surface $\Theta$ of the cloud. This region necessarily includes a portion of the initial surface $\Pi$, it might also contain all of it plus a portion of the center $\Gamma$, or even all of the center. In the last case, it occurs that some past-directed outgoing null ray arrives at the central singularity $\Sigma_0$ for some positive $j^*$, say. According to Theorem~\ref{Thm:CH} the subsequent outgoing light rays for $j < j^*$ must emanate from $\Sigma_0$ as well.
\item[{\bf Step 3}.] Similarly, we define a uniform grid $(0,R_k)$ along the singularity $\Sigma$, where $R_k = R_1 k/K$, $k=1,2,...,K$, and shoot from each of these gridpoints an outgoing radial light ray to the past. In order to perform the integration, we start at the point $(y,R) = (h, R_k + A h^4)$ close to the singularity, which comes from the truncation of the asymptotic expansion~(\ref {Eq:LightRaysSExp}). Then, we intersect the light ray with the relevant surfaces and determine $(U,V)$ at the intersection points like in the previous step. In this way, one obtains the image of the relevant surfaces beyond the past of $\Theta$.
\item[{\bf Step 4}.] Finally, we determine the conformal images of the singularities $\Sigma_0$ and $\Sigma$. By construction, the image of $\Sigma$ is a horizontal line in the $TX$ diagram with endpoints $(\pi/4,X_0)$, $(\pi/4,0)$, with $X_0 = (V_0 - U_0)/2$ defined below. The image of $\Sigma_0$, in turn, is defined by a unique value $V_0$ of $V$ and a whole range of values $U\in [U_0,U_1]$ for $U$. Indeed, according to Theorem~\ref{Thm:CH} there is a unique ingoing radial light ray terminating at $\Sigma_0$ while there are infinitely many outgoing radial light rays emanating from $\Sigma_0$, the earliest of which generates the Cauchy horizon. As a consequence, $\Sigma_0$ unfolds into a subset of the straight line $T + X = V_0$ in the conformal $TX$ diagram. Therefore, it is sufficient to compute $V_0$, $U_0$ and $U_1$ in order to determine its endpoints.

To compute $V_0$, we integrate the unique ingoing radial light ray to the past, starting  from the point $(y,R) = (1.25 h,\sqrt{4(1.25h)^3/(3\Lambda_0)})$ obtained from the truncation of the asymptotic expansion~(\ref{Eq:LightRaysS0Exp}), until it intersects $\Theta$ or $\Pi$. To compute $U_1$, we integrate to the future the outgoing radial light ray generating the Cauchy horizon starting from the point $(y,R) = (1.25 h,\sqrt{4(1.25h)^3/(3\Lambda_0)})$ until it intersects $\Theta$ or $\Sigma$. The intersection point determines the value for $U_1$. Finally, the value of $U_0$ is fixed by the requirement that $U_0 + V_0 = 2T = \pi/2$ since it must match the value for $U$ at the left endpoint of $\Sigma$.  
\end{enumerate}

\section{Numerical results}
\label{Sec:Results}

We apply our method for constructing the conformal diagram to the four-parameter family of initial data 
\begin{equation}
c(R) := c_0 \left[ 1 - \frac{6}{5} a\left( \frac{R}{R_1}\right)^2 
+ \frac{3}{7}(2a-1)\left(\frac{R}{R_1}\right)^4 \right],\qquad
q(R) := q_0 + q_1\left( \frac{R}{R_1} \right)^2,\qquad
0 \leq R \leq R_1,
\label{Eq:cqChoice}
\end{equation}
where the parameters $c_0$, $a$, $q_0$ and $q_1$ are subject to the inequalities $0 < a \leq 1$, $0 < R_1^2 c_0 < 3/2$, $0 < q_0 < 1$, and $0 < q_1 < 1-q_0$. These conditions guarantee that the corresponding initial density and velocity profiles $(\rho_0,v_0) = ( (R^3 c)'/(8\pi G R^2), -R\sqrt{(1-q^2)c})$ satisfy the assumptions (i)--(viii) on the interval $[0,R_1]$. Although the resulting density profile,
\begin{displaymath}
\rho_0(R) = \frac{3c_0}{8\pi G}\left[ 1 - 2a\left(\frac{R}{R_1}\right)^2
 + (2a-1)\left(\frac{R}{R_1}\right)^4 \right],\qquad
0 \leq R \leq R_1
\end{displaymath}
cannot be $C^\infty$-smoothly matched to zero at $R=R_1$, it still satisfies $\rho_0(R_1) = 0$ which implies that the metric is twice continuously differentiable across the surface of the cloud.

\subsection{Examples of numerically generated conformal diagrams}

\begin{figure}[h!]
\begin{center}
\includegraphics[width=14cm]{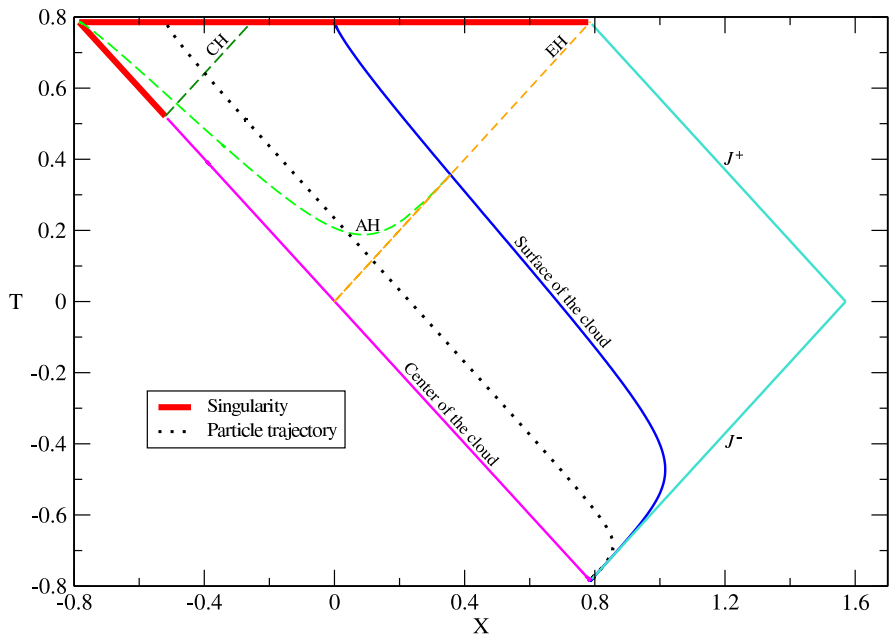}
\end{center}
\caption{\label{Fig:Hidden} Conformal diagram for the model described in equation~(\ref{Eq:cqChoice}) with the parameter choice $c_0 = 0.1691$, $a=0.3$, $q_0=0.75$ and $q_1=0.02$. The lines denoted by ``AH", ``EH" and ``CH" refer to the apparent, event and Cauchy horizons, respectively. In this case, the singularity is hidden inside the black hole region since all light rays emanating from it end at the spacelike singularity. Although the center of the cloud appears to be null in the diagram, closer inspection reveals that it is, in fact, time-like. The dotted line corresponds to the dust particle trajectory with initial areal radius $R_0 = 0.95R_1$. The diagram was generated with the step size $h = 0.0005$ and $1500$ and $500$ points, respectively, for the grids $y_j$ and $R_k$ (see steps 2 and 3 in the previous section).}
\end{figure}

\begin{figure}[h!]
\begin{center}
\includegraphics[width=12cm]{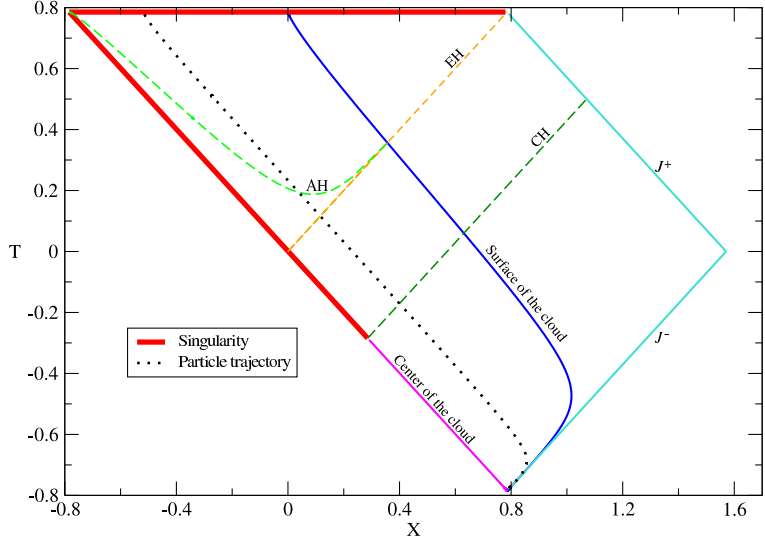}
\end{center}
\caption{\label{Fig:Naked_c} Conformal diagram for the model described in equation~(\ref{Eq:cqChoice}) with the same parameter choice as in the previous figure except that $c_0 = 0.1688$. In this case, there exists light rays emanating from the null part of the singularity which arrive at the surface of the cloud {\em earlier} than the apparent horizon. Therefore, a portion of the singularity is visible to distant observers {\em outside} the black hole region.}
\end{figure}

\begin{figure}[h!]
\begin{center}
\includegraphics[width=12cm]{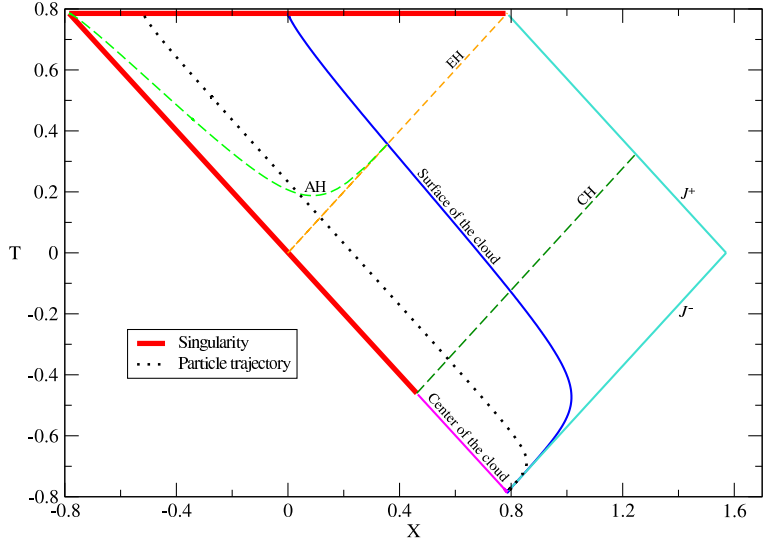}
\end{center}
\caption{\label{Fig:Naked_q} Conformal diagram for the model described in equation (\ref{Eq:cqChoice}) with the same parameter choice as in figure~\ref{Fig:Hidden}, except that $q_0 = 0.753$. Also in this case a portion of the singularity is visible {\em outside} the black hole region.}
\end{figure}

Conformal diagrams corresponding to three parameter choices are shown in figures~\ref{Fig:Hidden}, \ref{Fig:Naked_c} and \ref{Fig:Naked_q}. As anticipated in the previous section, the causal structure in the generic case is quite different than in the case of homogeneous, time-symmetric collapse illustrated in figure~\ref{Fig:Homogeneous}. First, we notice from the diagrams that the apparent horizon may be spacelike inside the cloud, whereas it is always timelike in the homogeneous, time-symmetric case. Next, while $\Sigma_0\cup\Sigma$ is spacelike in the latter, $\Sigma$ is still spacelike but the central singularity $\Sigma_0$ is null in the generic case, and therefore, it is visible to local observers. Moreover, the null part of the singularity may either be completely hidden inside the black hole region, as in figure~\ref{Fig:Hidden}, or a portion of it may be visible from future null infinity, as in figures~\ref{Fig:Naked_c} and \ref{Fig:Naked_q}.

It is worth stressing that no fine-tuning is required to construct the examples in which the naked singularity is globally visible. In fact, as Theorem~\ref{Thm:Naked} shows, it is sufficient to decrease the central density $\rho_0 = 3c_0/(8\pi G)$ to a small enough value in order to produce such singularities. For the initial data described in equation~(\ref{Eq:cqChoice}) with the parameter values used in figures~\ref{Fig:Hidden} and \ref{Fig:Naked_c} the bound~(\ref{Eq:c0Bound}) on $c_0$ in Theorem~\ref{Thm:Naked} gives
\begin{displaymath}
\mu^2\le 0.062,
\end{displaymath}
which is consistent with the numerical results in those figures. However, the numerical results also indicate that our bound is far from optimal, since the transition from local to global visibility occurs around $c_0\approx 0.169$ which is much larger than $\mu^2$.

\begin{figure}[h!]
\begin{center}
\includegraphics[width=12cm]{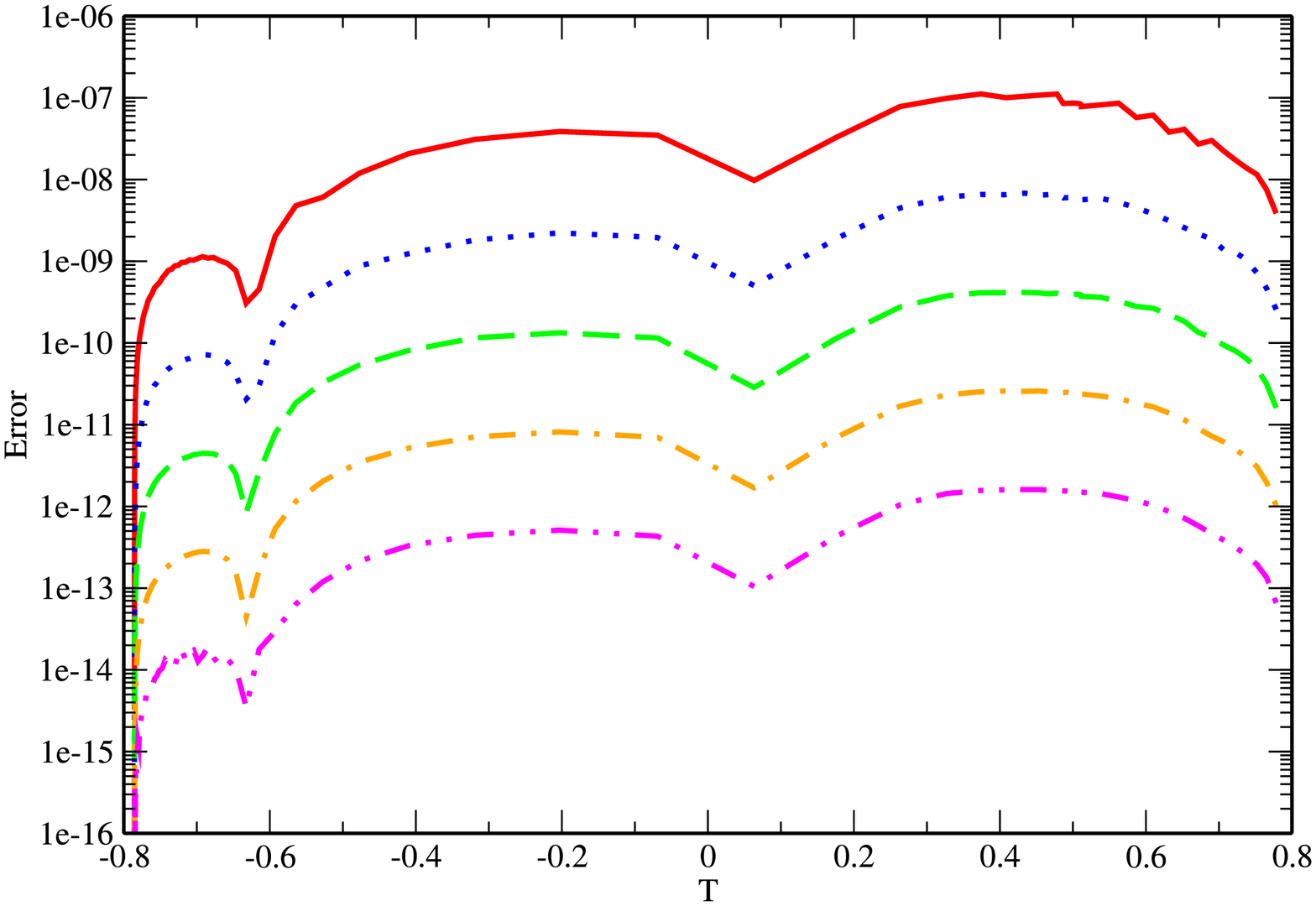}
\end{center}
\caption{\label{Fig:Tr_conv} Numerical error for points on the particle trajectory shown in figure~\ref{Fig:Naked_c}. The continuous line gives the Euclidean distance between corresponding points computed with step sizes $h=0.01$ and $h/2$. The dotted one represents the error for points computed with step sizes $h/2$ and $h/4$, and so forth.}
\end{figure}

We have verified the self-convergence of our numerical results by generating the conformal diagram using different step sizes $h$. In table~\ref{Tab:Convergence} we show the values for the key quantities $U_1$ and $V_0$ in the example of figure~\ref{Fig:Naked_c}, which determine the location of the earliest singular point in the collapse, from which the Cauchy horizon emanates. The results show self-convergence to an order between three and four.
\begin{table*}[!t]\centering
\begin{tabular}{|c|c|c|c|c || c|c|c|}
\hline
Resolution & Step size & $U_1$ & Error & CF & $V_0$ & Error& CF\\
\hline
1 &$h=0.01$&$-5.764905579201557E-01$ &            ---            &              ---             & $1.120836074650528E-03$ &            ---              &     --- \\
2 &$h/2$    &$-5.709420746741624E-01$ & $5.54848E-3$ &              ---             & $1.120831798152899E-03$ & $4.2765E-9$   &     --- \\
3 &$h/4$    &$-5.704917279004125E-01$ & $4.50347E-4$ & $12.3205$ & $1.120831389468409E-03$ & $4.08684E-10$   & $10.4641$ \\
4 &$h/8$    &$-5.704566274816609E-01$ & $3.51004E-5$ & $12.8302$ & $1.120831350043764E-03$ & $3.94246E-11$ & $10.3662$ \\
5 &$h/16$  &$-5.704540486177565E-01$ & $2.57886E-6$ & $13.6108$ & $1.120831346181979E-03$ & $3.86179E-12$ & $10.2089$ \\
6 &$h/32$  &$-5.704538804441139E-01$ & $1.68174E-7$ & $15.3345$ & $1.120831345796034E-03$ & $3.85945E-13$ & $10.0061$\\
\hline
\end{tabular}
\caption{\label{Tab:Convergence} Self-convergence test for the quantities $U_1$ and $V_0$ in the example shown in figure~\ref{Fig:Naked_c}. For the $i-$th resolution the error is defined as $E_i := |U_i - U_{i-1}|$, and the convergence factor as $CF_i := E_{i-1}/E_{i}$.}
\end{table*}
A similar self-convergence test is performed for the dust particle trajectory shown in figure~\ref{Fig:Naked_c}, corresponding to an initial areal radius of $R_0 = 0.95 R_1$. In  figure~\ref{Fig:Tr_conv}, we show the numerical error for different points on this trajectory. The numerical error is estimated by computing the Euclidean norm between two successive resolutions ($h$ and $h/2$, $h/2$ and $h/4$ and so forth), and clearly decreases as resolution is increased. In order to quantify these errors, we show in table~\ref{Tab:Tr_conv_factor} for each fixed resolution the maximum of this error with respect to all points on the trajectory. The results exhibit fourth order self-convergence.

\begin{table*}[!t]\centering
\begin{tabular}{|c|c|c|c|}
\hline
Resolution &Step size & Maximum error & CF\\
\hline
1 &$h=0.01$&                        ---                        &          ---        \\
2 &$h/2$       &$1.115221660178996E-7  $ &         ---         \\
3 &$h/4$       &$6.812854288367279E-9  $ & $16.3693$ \\
4 &$h/8$       &$4.148846069813869E-10$ & $16.4211$ \\
5 &$h/16$     &$2.588085455658281E-11$ & $16.0306$ \\
6 &$h/32$     &$1.608422340951133E-12$ & $16.0908$ \\
\hline
\end{tabular}
\caption{\label{Tab:Tr_conv_factor} Self-convergence test for the dust particle trajectory shown in figure~\ref{Fig:Naked_c}. The $i$th maximum error is defined as the largest Euclidean distance between corresponding points on this trajectory computed for resolutions $i$ and $i-1$. The convergence factor is computed as in table~\ref{Tab:Convergence}. Notice that a step size of order $h = 0.005$ is sufficient to keep the maximum error below $1.2\times 10^{-7}$.}
\end{table*}

\subsection{Phase space diagrams}
\label{Sec:PhaseSpace}

\begin{figure}[h!]
\begin{center}
\includegraphics[width=12cm]{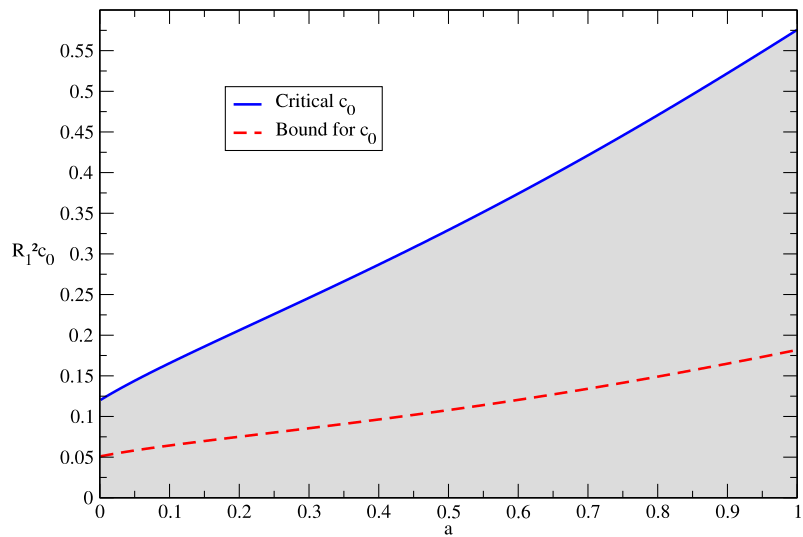}
\end{center}
\caption{\label{Fig:c0bound} A cut through the phase diagram corresponding to the four-parameter family of initial data given in equation~(\ref{Eq:cqChoice}). The cut corresponds to the subset with fixed parameters $q_0 = 0.75$ and $q_1 = 0.02$. The shaded region corresponds to initial data giving rise to a naked singularity which is globally visible. The dashed line describes the upper bound for $c_0$ from Theorem~~\ref{Thm:Naked}.}
\end{figure}

As an application of our results, we explore some features of the phase diagram corresponding to the four-parameter family of initial data given in equation~(\ref{Eq:cqChoice}). In figure~\ref{Fig:c0bound} we show a subset for fixed $q$ of this diagram, and the critical line in the $a$-$c_0$-plane that divides the regions corresponding to initial data giving rise to black holes and globally naked singularities, respectively.  From the figure, it can be observed that density profiles which are nearly flat close to the center, corresponding to $a\ll 1$, have a lower critical value for $c_0$ than density profiles which are concentrated near the center. In this sense, diluted profiles favor the formation of black holes while concentrated profiles favor the formation of globally naked singularities. Notice also that although the upper bound for $c_0$ obtained in Theorem~\ref{Thm:Naked} is far from optimal, it nevertheless describes the correct qualitative behavior for the critical line.

\begin{figure}[h!]
\begin{center}
\includegraphics[width=12cm]{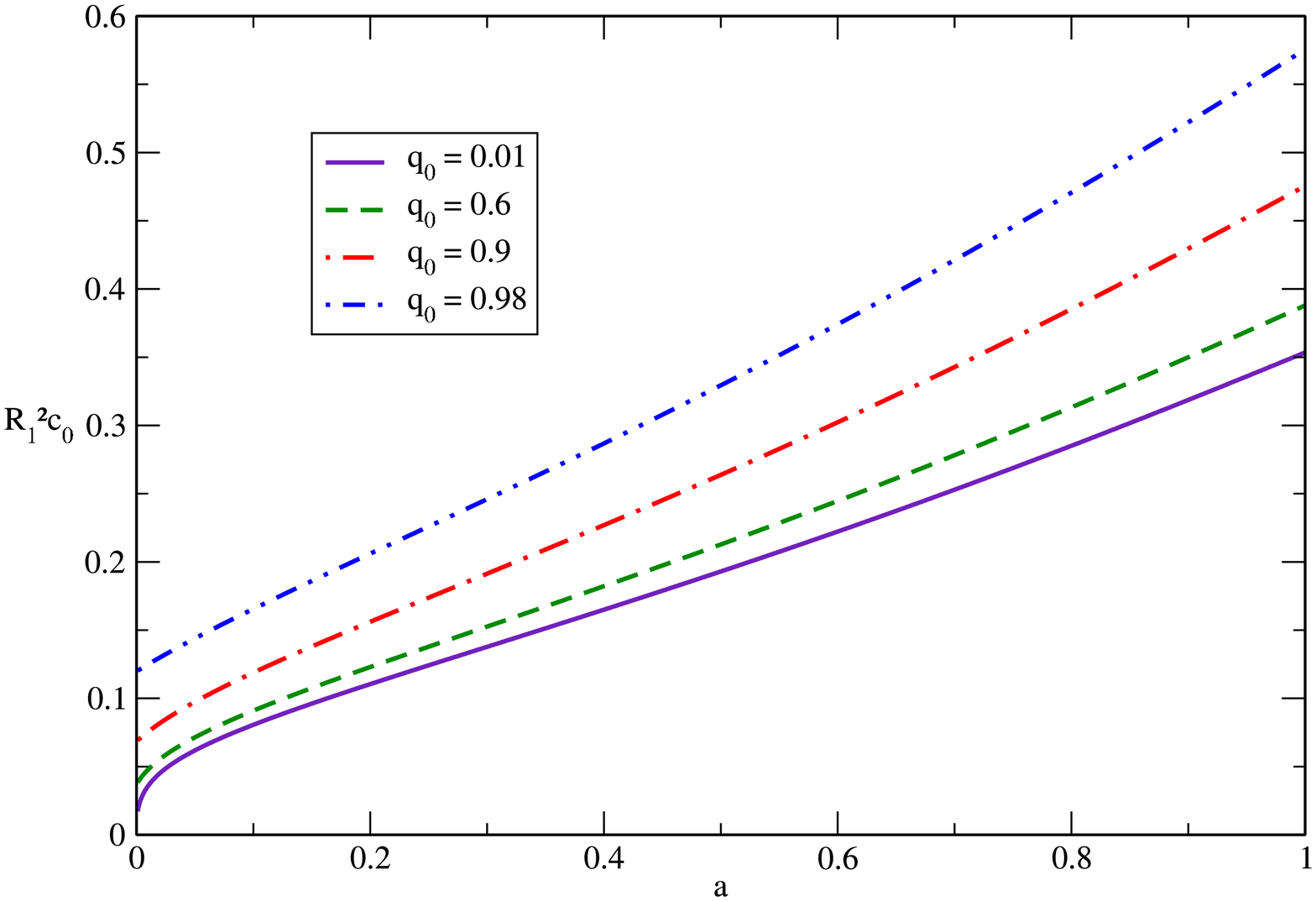}
\end{center}
\caption{\label{Fig:qbound} The critical line, dividing the regions of initial data giving rise to black holes and globally naked singularities, respectively, for different values of $q_0$ and $q_1 = 0.01$.}
\end{figure}

In figure~\ref{Fig:qbound} we show the critical line in the $a$-$c_0$-plane for different values of $q_0$ and $q_1 = 0.01$. As we see, the critical value for $c_0$ decreases as $q_0$ decreases from $0.98$ to $0.01$. Since $1 - q^2$ is the ratio between the initial kinetic energy and the magnitude of the initial potential energy, this means that large initial velocities in the negative radial direction favor the formation of black holes.

\section{Conclusions}
\label{Sec:Conclusions}

We have presented a numerical method for constructing the conformal diagram inside a spherically symmetric dust cloud which undergoes complete gravitational collapse. The construction is based on the integration of radial null geodesics and a careful analysis of their behavior close to the singularity. Our assumptions on the initial data, namely the initial density and velocity profiles, should be physically reasonable and generic within the limitation of spherical symmetry and zero pressure. Confirming previously known results~\cite{dC84,rN86,pJiD93,Joshi-Book}, we find that under these assumptions, the shell-focusing singularity always consists of a space-like part, which matches the one of the Schwarzschild exterior, and a null part which is "naked" in the sense that it is visible, at least to observers which are sufficiently close to the center of the cloud.

Based on our method, we analyzed a four-parameter family of initial data and determined under which circumstances the null part of the shell-focusing singularity forms sufficiently early such that a portion of it lies outside the event horizon, resulting in a naked singularity which is globally visible. We find that this can be arranged fairly simply without fine-tuning the data. This shows that it is not difficult to create naked singularities which are in causal contact with future null infinity, confirming the results in~\cite{dC84} for time-symmetric initial data. More importantly, however, these findings also indicate that the formation of a globally naked singularity is stable with respect to perturbations within the spherical symmetric, zero pressure model. In fact, this stability statement is confirmed and strengthened by Theorem~\ref{Thm:Naked}, which provides a new bound on the initial data guaranteeing that the corresponding spacetime contains such a singularity. The corresponding spacetimes are not globally hyperbolic, and as a consequence, it is not possible to determine the evolution of test fields obeying hyperbolic partial differential equations with light speed propagation beyond the Cauchy horizon without imposing boundary conditions at the null singularity. Since the Cauchy horizon is located outside the black hole region, this implies, in particular, that it is not even possible to predict the evolution of such test fields in the Schwarzschild exterior for $r > 2m$.

Besides the results already mentioned, there are several possible applications of our numerical algorithm. First, it provides a tool to systematically analyze the phase space of spherically symmetric dust collapse, identifying the class of initial data leading to globally naked singularities. As an example, we have explored in section~\ref{Sec:PhaseSpace} different cuts through the phase space, and determined the critical set within these cuts which separates the black holes from the globally naked singularities. It should be interesting to perform a more exhaustive analysis, including, for example, initial data with arbitrarily concentrated density profiles or families of initial data which contain the Oppenheimer-Snyder model as a limiting case. Second, the numerical construction of the conformal coordinates may be useful for describing the propagation of test fields on the collapsing spacetimes considered here. For example, the wave equation $\Box\Phi = 0$ reduces to a family of flat space wave equations of the simple form
\begin{equation}
\frac{\partial^2\psi_{\ell m}}{\partial T^2} - \frac{\partial^2\psi_{\ell m}}{\partial X^2}
 + V(T,X)\psi_{\ell m} = 0,\qquad
 \Phi(T,X,\vartheta,\varphi) 
 = \frac{1}{r}\sum\limits_{\ell m} \psi_{\ell m}(T,X) Y^{\ell m}(\vartheta,\varphi)
\end{equation}
with a potential $V$, when decomposed into spherical harmonics $Y^{\ell m}$. Therefore, properties of the solutions outside the Cauchy horizon can, in principle, be inferred from the structure of the potential $V$.

It remains to be seen whether or not our results can be extrapolated to the realistic gravitational collapse, in which pressure and angular momentum may delay the formation of the singularity. If these effects are sufficiently strong, it is imaginable that the resulting singularity is hidden inside the black hole, such that weak cosmic censorship is valid. Unfortunately, analyzing scenarios with more realistic equations of state in the absence of symmetries requires much more efforts since exact solutions are not available anymore, at least not for generic initial data. Intermediate steps toward understanding the general case are the following: first, it should be possible to generalize our method to the case of a spherically symmetric collapse with nonzero pressure, since it is based on the numerical integration of radial null rays and their asymptotic behavior near the shell-focusing singularity. Next, one could consider a collapsing fluid star which is slightly nonspherical, in which case the problem may be analyzed using perturbation theory for spherically symmetric spacetimes. In the dust case, numerical work~\cite{hItHkN00} and recent analytic results~\cite{eDbN11} for the self-similar collapse show that linear fluctuations from smooth initial data grow without bound when approaching the Cauchy horizon, providing evidence that the latter is unstable at the linearized level. Finally, the problem can be approached using numerical approximation. In~\cite{sSsT91}, a numerical code was used to evolve collisionless prolate gas spheroids, and it was shown that when sufficiently large, these spheroids form a curvature singularity before an apparent horizon appears. However, as pointed out in~\cite{rWvI91}, this does not exclude the formation of trapped surfaces in the maximally extended spacetime, and so the numerical work in~\cite{sSsT91} does not imply a violation of cosmic censorship at this stage. More recently, the gravitational collapse has also been analyzed in five~\cite{lLfP10,yYhS11} and higher-dimensional spacetimes~\cite{rGpJ04,rGpJ07}.

Even if it turned out that generically, singularities in the nonspherical gravitational collapse with a realistic equation of state are hidden inside black holes, it would still be interesting to understand what happens in the limit when the collapse becomes spherical and pressure can be neglected. Presumably, the fact that in this limit naked singularities which are visible from future null infinity exist should have an imprint on the perturbed case.


\acknowledgments

We thank Thomas Zannias for fruitful and stimulating discussions. This work was supported in part by Grants CONACyT 46521 and 61173 and CIC 4.19 to Universidad Michoacana.

\appendix
\section{An existence theorem for nonlinear perturbations of regular singular points}
\label{App:Theorem}

In this appendix, we include a short proof of the following theorem which we used in section~\ref{Sec:Theorems} to show the local existence of light rays emanating or terminating at the singularity. It is based on basic tools from the theory of dynamical systems.\footnote{We thank Thomas Zannias for pointing out to us the elegant proof of this theorem.}  For a generalization to systems of equations and applications to relativistic stars, see Ref.~\cite{aRbS91}.

\begin{theorem}
\label{Thm:RegSingPert}
Let $\beta > 0$, and let $D\subset \Real^2$ be an open subset of $\Real^2$ which contains the origin. Furthermore, let $f: D \to \Real$ be a $C^\infty$-function. Then, the differential equation
\begin{equation}
x\frac{dy}{dx} + \beta y = x f(x,y)
\label{Eq:RegularSingularPert}
\end{equation}
has a unique local $C^\infty$-solution $y: (0,\varepsilon) \to \Real$ which is bounded.
Moreover, this solution satisfies
\begin{displaymath}
\lim\limits_{x\to 0} y(x) = 0,\qquad
\lim\limits_{x\to 0}\frac{dy}{dx}(x) = \frac{f(0,0)}{\beta+1}.
\end{displaymath}
\end{theorem}

\proof Define $\alpha := f(0,0)$, and introduce the parameter $t = -\log(x)$ for $x > 0$. Then, the solutions of equation~(\ref{Eq:RegularSingularPert}) are given by the trajectories of the autonomous system
\begin{equation}
\frac{d}{dt} u = A u + F(u),
\label{Eq:Dynamical}
\end{equation}
with
\begin{equation}
u = \left( \begin{array}{c} x \\ y \end{array} \right),\qquad
A = \left( \begin{array}{rr} -1 & 0 \\ -\alpha & \beta \end{array} \right),\qquad
F(u) = \left( \begin{array}{c} 0 \\ x\left[ f(0,0) - f(x,y) \right] \end{array} \right).
\end{equation}
The dynamical system described by equation~(\ref{Eq:Dynamical}) has a stationary point at $u=0$, and the linearized system at this point is given by the matrix $A$. The eigenvalues of $A$ are $-1$ and $\beta > 0$, with corresponding one-dimensional stable and unstable manifolds describing the sets of points in phase space converging to $u=0$ for $t\to\infty$ and $t\to -\infty$, respectively, see for example Ref.~\cite{Hartman-Book}. Therefore, the solutions of equation~(\ref{Eq:RegularSingularPert}) which are bounded for small $x = \exp(-t)$ correspond to the stable manifold. Its tangent vector at $u=0$ is given by the eigenspace of $A$ corresponding to the eigenvalue $-1$. Since this eigenspace is generated by the vector $(\beta+1,\alpha)$, the slope of the solution at $u=0$ is $\alpha/(\beta+1)$.
\qed

\section{Proof of Lemma~\ref{Lemma:uniqueness}}
\label{App:Lemma}

Here we prove Lemma~\ref{Lemma:uniqueness} which is a statement about local, $C^1$-solutions $\varphi$ of equation~(\ref{Eq:dphi/du2/3}) satisfying $\lim\limits_{u\to 0}\varphi(u) = 0$. For this, it is convenient to rewrite equation~(\ref{Eq:dphi/du2/3}) in the form
\begin{equation}
\frac{dy}{du} = A(y,u)\left( \frac{u}{y} \right)^2
\left[ 1 - \epsilon B(y,u)\sqrt{u}\left( \frac{u}{y} \right) \right]
 - \frac{3\epsilon}{4} B(y,u)\sqrt{u},
\label{Eq:dphi/du2/3Bis}
\end{equation}
where the functions $A,B: [0,1)\times [0,1]\to\Real$ are defined as
\begin{eqnarray*}
A(y,u) &:=& \left. \frac{3}{4}R_1^2 \Lambda(y,R) \sqrt{1-q(R)^2y^2} 
 \right|_{R = R_1 u^{3/2}}, \\
B(y,u) &:=& \left.  R_1 Q(R)\sqrt{1 - q(R)^2 y^2}  \right|_{R = R_1 u^{3/2}}.
\end{eqnarray*}
According to Lemma~\ref{Lem:Elementary}, these functions are continuous, and they satisfy
\begin{eqnarray*}
A_0 &:=& A(0,0) = \frac{3}{4} R_1^2\Lambda(0,0) = \lambda^3 > 0,\\
B_0 &:=& B(0,0) = R_1 Q_0 > 0, 
\end{eqnarray*}
where $\lambda$ is defined in equation~(\ref{Eq:lambda}) and $Q_0 := Q(0) > 0$. Therefore, given $\delta > 0$ with $\delta < \min\{ A_0,B_0 \}$, there exists $\delta_y > 0$ and $\delta_u > 0$ small enough such that $D:=[0,\delta_y]\times [0,\delta_u]\subset [0,1)\times [0,1]$ and
\begin{eqnarray*}
0 < A_0 - \delta \leq &A(y,u)& \leq A_0 + \delta,\\
0 < B_0 - \delta \leq &B(y,u)& \leq B_0 + \delta
\end{eqnarray*}
for all $(y,u)\in D$. Let $\varphi: (0,\delta_u) \to (0,\delta_y)$ be a local solution of equation~(\ref{Eq:dphi/du2/3Bis}) such that $\lim\limits_{u\to 0}\varphi(u) = 0$, and set
\begin{displaymath}
m := \inf\limits_{0 < u < \delta_u} \frac{\varphi(u)}{u} \geq 0,\qquad
M := \sup\limits_{0 < u < \delta_u} \frac{\varphi(u)}{u} \leq \infty.
\end{displaymath}
We are now ready to prove the lemma.

\begin{enumerate}
\item[(i)] Suppose $m > 0$. Then, we can use $u/y\leq 1/m$ in equation~(\ref{Eq:dphi/du2/3Bis}) and estimate
\begin{displaymath}
\frac{dy}{du} \leq \frac{A_0 + \delta}{m^2}
\left[ 1 + \frac{B_0 + \delta}{m}\sqrt{\delta_u} \right]
 + \frac{3(B_0+ \delta)}{4}\sqrt{\delta_u} =: M',
\end{displaymath}
for $0 < u < \delta_u$. Since $\lim\limits_{u\to 0}\varphi(u) = 0$ this implies that $\varphi(u)\leq M' u$ for all $0 < u < \delta_u$. Therefore, $M\leq M' < \infty$.
\item[(ii)] Conversely, suppose $M < \infty$, and assume $\epsilon=-1$. Using the estimate $u/y \geq 1/M$ in equation~(\ref{Eq:dphi/du2/3Bis}) and the positivity of $B$ on $D$, we obtain
\begin{displaymath}
\frac{dy}{du} \geq \frac{A_0 - \delta}{M^2} =: m' > 0,
\end{displaymath}
which implies $\varphi(u)\geq m' u$ for all $0 < u < \delta_u$. Therefore, $m\geq m' > 0$.
\item[(iii)] Finally, suppose $m > 0$ and $M < \infty$. According to the L'H\^opital's rule (see, for instance, Ref.~\cite{Spivak-calculus-Book}), we have
\begin{displaymath}
\lim\limits_{u\to 0} \frac{\varphi(u)^3}{u^3} 
 = \lim\limits_{u\to 0}\left(\frac{\varphi(u)}{u}\right)^2 \frac{d\varphi}{du} 
 = A_0 = \lambda^3,
\end{displaymath}
where we have used equation~(\ref{Eq:dphi/du2/3Bis}) and the boundedness of $u/y$ in the second step. Therefore, $\lim\limits_{u\to 0} \varphi(u)/u = \lambda$.
\end{enumerate}
\qed

\section{The surface of the cloud in Kruskal coordinates}
\label{App:Kruskal}

The purpose of this appendix is to construct Penrose-Kruskal coordinates along the surface of the cloud, $\Theta$. For this, we start with the expressions for the Kruskal null coordinates which are related to the standard Schwarzschild coordinates $(t,r)$ by~\cite{Wald-Book}:
\begin{eqnarray*}
U &=& \pm\sqrt{\left| \frac{r}{2m} - 1\right|}\exp \left( \frac{r-t}{4m} \right),\\
V &=& \sqrt{\left| \frac{r}{2m} - 1 \right|}\exp \left( \frac{r+t}{4m} \right),
\end{eqnarray*}
where the choice of sign ($+/-$) corresponds to the region inside or outside the event horizon, respectively. These coordinates satisfy the relation
\begin{equation}
UV = \left(1 - \frac{r}{2m} \right) \exp \left(\frac{r}{2m} \right),\qquad
r > 0.
\label{Eq:UV}
\end{equation}
The surface $\Theta$ is generated by the trajectories of freely falling particles with zero angular momentum in the Schwarzschild spacetime, for which the equations of motion are given by equation~(\ref{Eq:1DMechanical}) with $R=R_1$. The coordinate $t$ along the geodesic is determined by the conservation of energy equation
\begin{displaymath}
\frac{dt}{d\tau} = \frac{\sqrt{1 + 2E_1}}{1 - \frac{2m_1}{r}},
\end{displaymath}
where $E_1 := 2E(R_1)$ and $m_1 := m(R_1)$. Using this, equation~(\ref{Eq:1DMechanical}) and the definition of $U$ and $V$, we find the following equations,
\begin{eqnarray}
4m_1 \frac{d}{dr}\log(U) &=& \frac{r}{r-2m_1}  \left( 1 + \sqrt{ \frac{1+2E_1} {\frac{2m_1}{r} +2E_1} }\right),\label{Eq:dU/dr}\\
4m_1 \frac{d}{dr}\log(V) &=& \frac{r}{r-2m_1}  \left( 1 - \sqrt{ \frac{1+2E_1} {\frac{2m_1}{r} +2E_1} }\right)\label{Eq:dV/dr},
\end{eqnarray}
which are valid both outside and inside the event horizon. Integrating equation~(\ref{Eq:dU/dr}), we obtain
\begin{displaymath}
U(y) = \frac{1}{a_1} \left( \sqrt{a_1^2-b_1^2y^2} - y\sqrt{1-b_1^2} \right) 
\exp \left\{ \frac{y^2}{2a_1^2} - \frac{\sqrt{1-b_1^2}}{2b_1^2}
\left[ \frac{1+2b_1^2}{b_1} \arctan \left( \frac{\sqrt{a_1^2-b_1^2y^2}}{b_1y} \right)
 +  \frac{y}{a_1^2}\sqrt{a_1^2-b_1^2y^2}    \right] \right\},
\end{displaymath}
where $a_1^2 := 2m_1/R$, $b_1^2:=-2E_1$ and $y^2 = r/R_1$. From the relation~(\ref{Eq:UV}) it follows immediately that
\begin{displaymath}
V(y) = \frac{1}{a_1} \left( \sqrt{a_1^2-b_1^2y^2} + y\sqrt{1-b_1^2} \right)
\exp \left\{ \frac{y^2}{2a_1^2} + \frac{\sqrt{1-b_1^2}}{2b_1^2}
\left[ \frac{1+2b_1^2}{b_1} \arctan \left( \frac{\sqrt{a_1^2-b_1^2y^2}}{b_1y} \right)
 +  \frac{y}{a_1^2}\sqrt{a_1^2-b_1^2y^2}    \right] \right\}.
\end{displaymath}

A translation in $t$ induces the transformations $U\mapsto \kappa U$ and $V\mapsto V/\kappa$, with $\kappa > 0$. We choose $\kappa$ such that $U(0) = V(0) = 1$. The Penrose-Kruskal coordinates $U_\Theta$ and $V_\Theta$ in section~\ref{Sec:Diagrams} are obtained from the resulting rescaled quantities after compactification, $U_\Theta(y) := \arctan( U(y) )$, $V_\Theta(y) := \arctan( V(y) )$. 

Using equations~ (\ref{Eq:dU/dr}) and~(\ref{Eq:dV/dr}) and the fact that $\dot{r} < 0$ along the surface, it is not difficult to check that $\dot{U}$ and $\dot{V}$ are positive everywhere along $\Theta$. Therefore, the conformal factor in section~\ref{Sec:Diagrams}, $\Omega = 1/\sqrt{\dot{U}_\Theta\dot{V}_\Theta}$, is well defined.

\bibliographystyle{unsrt}
\bibliography{../References/refs_collapse}

\begin{thebibliography}{10}

\bibitem{rSsY83}
R.~Schoen and S.-T. Yau.
\newblock The existence of a black hole due to condensation of matter.
\newblock {\em Comm.Math.Phys.}, 90:575--579, 1983.

\bibitem{HawkingEllis-Book}
S.W. Hawking and G.F.R. Ellis.
\newblock {\em The Large Scale Structure of Space Time}.
\newblock Cambridge University Press, Cambridge, 1973.

\bibitem{Wald-Book}
R.M. Wald.
\newblock {\em General Relativity}.
\newblock The University of Chicago Press, Chicago, London, 1984.

\bibitem{rP69}
R.~Penrose.
\newblock Gravitational collapse: The role of general relativity.
\newblock {\em Riv. del Nuovo Cimento}, 1:252--276, 1969.

\bibitem{rW97}
R.M. Wald.
\newblock Gravitational collapse and cosmic censorship.
\newblock {\em arXiv:gr-qc/9710068v3}, 1997.

\bibitem{pJ00}
P.S. Joshi.
\newblock Gravitational collapse: The story so far.
\newblock {\em Pramana Journal of Physics}, 55:529--544, 2000.

\bibitem{pYhShM73}
P.~Yodzis, H.-J. Seifert, and H.~M\"uller zum Hagen.
\newblock On the occurrence of naked singularities in general relativity.
\newblock {\em Comm. Math. Phys.}, 34:135--148, 1973.

\bibitem{dElS79}
D.M. Eardley and L.~Smarr.
\newblock Time functions in numerical relativity: Marginally bound dust
  collapse.
\newblock {\em Phys. Rev. D}, 19:2239--2259, 1979.

\bibitem{dC84}
D.~Christodoulou.
\newblock Violation of cosmic censorship in the gravitational collapse of a
  dust cloud.
\newblock {\em Comm. Math. Phys.}, 93:171--195, 1984.

\bibitem{rN86}
R.P.A.C Newman.
\newblock Strengths of naked singularities in {T}olman-{B}ondi spacetimes.
\newblock {\em Class. Quantum Grav.}, 3:527--539, 1986.

\bibitem{pJiD93}
P.S. Joshi and I.H. Dwivedi.
\newblock Naked singularities in spherically symmetric inhomogeneous
  {T}olman-{B}ondi dust cloud collapse.
\newblock {\em Phys. Rev. D}, 47:5357--5369, 1993.

\bibitem{Joshi-Book}
P.S. Joshi.
\newblock {\em Gravitational Collapse and Spacetime Singularities}.
\newblock Cambridge University Press, Cambridge, 2008.

\bibitem{MTW-Book}
C.W. Misner, K.S. Thorne, and J.A. Wheeler.
\newblock {\em Gravitation}.
\newblock W. H. Freeman, 1973.

\bibitem{cMdS64}
C.W. Misner and D.H. Sharp.
\newblock Relativistic equations for adiabatic, spherically symmetric
  gravitational collapse.
\newblock {\em Phys. Rev.}, 136:B571--B576, 1964.

\bibitem{jOhS39}
J.R. Oppenheimer and H.~Snyder.
\newblock On continued gravitational contraction.
\newblock {\em Phys. Rev.}, 56:455--459, 1939.

\bibitem{pSaL99}
P.~Szekeres and A.~Lun.
\newblock What is a shell-crossing singularity?
\newblock {\em J. Austral. Math. Soc. Ser. B}, 41:167--179, 1999.

\bibitem{Straumann-Book}
N.~Straumann.
\newblock {\em General Relativity and Relativistic Astrophysics}.
\newblock Sprin\-ger-Verlag, Berlin, 1984.

\bibitem{sDpJiD02}
S.S. Deshingkar, P.S. Joshi, and I.H. Dwivedi.
\newblock Appearance of the central singularity in spherical collapse.
\newblock {\em Phys. Rev. D}, 65:084009, 2002.

\bibitem{tSpJ96}
T.P. Singh and P.S. Joshi.
\newblock The final fate of spherical inhomogeneous dust collapse.
\newblock {\em Class. Quantum Grav.}, 13:559--571, 1996.

\bibitem{bNfM01}
F.C. Mena and B.C. Nolan.
\newblock Non-radial null geodesics in spherical dust collapse.
\newblock {\em Class. Quantum Grav.}, 18:4531--4548, 2001.

\bibitem{Recipies-Book}
W.H. Press, S.A. Teukolsky, W.T. Vetterling, and B.P. Flannery.
\newblock {\em Numerical Recipes in Fortran}.
\newblock Cambridge University Press, Cambridge, 1992.

\bibitem{nOoS10}
N.~Ortiz and O.~Sarbach.
\newblock Conformal diagrams for the gravitational collapse of a spherically
  symmetric dust cloud.
\newblock {\em AIP Conf. Proc.}, 1256:349--356, 2010.
\newblock Proceedings of VIII Mexican School on Gravitation and Mathematical
  Physics.

\bibitem{hItHkN00}
H.~Iguchi, T.~Harada, and K.~Nakao.
\newblock Gravitational radiation from a naked singularity. 2. {E}ven parity
  perturbation.
\newblock {\em Prog. Theor. Phys.}, 103:53--72, 2000.

\bibitem{eDbN11}
E.M. Duffy and B.C. Nolan.
\newblock Cosmic censorship for self-similar spherical dust collapse.
\newblock 2011.
\newblock arXiv:1108.1103 [gr-qc].

\bibitem{sSsT91}
S.L. Shapiro and S.A. Teukolsky.
\newblock Formation of naked singularities: The violation of cosmic censorship.
\newblock {\em Phys. Rev. Lett.}, 66:994--997, 1991.

\bibitem{rWvI91}
R.M. Wald and V.~Iyer.
\newblock Trapped sufaces in the {S}chwarzschild geometry and cosmic
  censorship.
\newblock {\em Phys. Rev. D}, 44:R3719--R3722, 1991.

\bibitem{lLfP10}
L.~Lehner and F.~Pretorius.
\newblock Black strings, low viscosity fluids, and violation of cosmic
  censorship.
\newblock {\em Phys.Rev.Lett.}, 105:101102, 2010.

\bibitem{yYhS11}
Y.~Yamada and H.~Shinkai.
\newblock Formation of naked singularities in five-dimensional space-time.
\newblock {\em Phys.Rev. D}, 83:064006, 2011.

\bibitem{rGpJ04}
R.~Goswami and P.S. Joshi.
\newblock Cosmic censorship in higher dimensions.
\newblock {\em Phys. Rev. D}, 69:104002, 2004.

\bibitem{rGpJ07}
R.~Goswami and P.S. Joshi.
\newblock Spherical gravitational collapse in {$N$} dimensions.
\newblock {\em Phys. Rev. D}, 76:084026, 2007.

\bibitem{aRbS91}
A.D. Rendall and B.G. Schmidt.
\newblock Existence and properties of spherically symmetric static fluid bodies
  with a given equation of state.
\newblock {\em Class. Quantum Grav.}, 8:985--1000, 1991.

\bibitem{Hartman-Book}
P.~Hartman.
\newblock {\em Ordinary Differential Equations}.
\newblock John Wiley \& Sons, Inc., New York, 1964.

\bibitem{Spivak-calculus-Book}
M.~Spivak.
\newblock {\em Calculus}.
\newblock W. A. Benjamin, Inc., New York, 2001.

\end{thebibliography}

\end{document}